\documentclass[%
groupedaddress,
preprint,
 amsmath,amssymb,
 aps,
]{revtex4-2}

\usepackage{longtable}
\usepackage{cancel}
\usepackage{graphicx}
\usepackage{dcolumn}
\usepackage{bm}
\usepackage[overleftarrow, old-arrows]{overarrows}

\usepackage{booktabs}
\usepackage{tabularx}
\usepackage{makecell}
\usepackage{hhline}
\usepackage{adjustbox}
\usepackage{enumitem}
\usepackage{xcolor}
\usepackage{braket}
\usepackage{amsthm}
\usepackage[colorlinks=true,allcolors=blue]{hyperref}  

\renewcommand{\vec}[1]{\boldsymbol{#1}}   
\newcommand{\mat}[1]{\mathbf{#1}}         
\newcommand{\smat}[1]{\mathrm{#1}}        

\newcommand{\bbR}{\mathbb{R}}             
\newcommand{\bbC}{\mathbb{C}}             
\newcommand{\bbH}{\mathbb{H}}             

\newcommand{\qI}{1}                       
\newcommand{\qi}{\check{\imath}}          
\newcommand{\qj}{\check{\jmath}}          
\newcommand{\qk}{\check{k}}               
\newcommand{\qv}{\check{\boldsymbol{q}}}  
\newcommand{\qdot}{\cdot}

\newcommand{\qMb}[2]{{^{\text{#1}\!}\mat{M}}_{#2}}
\newcommand{\qNb}[2]{{^{\text{#1}\!}\mat{N}}_{#2}}
\newcommand{\qQb}[2]{{^{\text{#1}\!}\mat{Q}}_{#2}}

\begin{document}

\title{Biquaternion Algebra with Bilinear Multiplication: A General, Elegant, and Computationally Advantageous Framework for Relativistic Electronic Structure Calculations on CPUs and GPUs
}

\author{Sylvia Kaviraj}
\affiliation{Department of Physical and Theoretical Chemistry, Faculty of Natural Sciences, Comenius University in Bratislava, Slovakia}

\author{Stanislav Komorovsky}
\affiliation{Institute of Inorganic Chemistry, Slovak Academy of Sciences, Bratislava, Slovakia}

\author{Trond Saue}
\affiliation{Laboratoire de Chimie et Physique Quantiques, UMR 5626 CNRS --- Université de Toulouse, France}

\author{Michal Repisky}
\affiliation{Department of Physical and Theoretical Chemistry, Faculty of Natural Sciences, Comenius University in Bratislava, Slovakia}
\affiliation{Hylleraas Centre for Quantum Molecular Sciences, Faculty of Science and Technology, UiT - The Arctic University of Norway, Tromsø, Norway}
\email{michal.repisky@uit.no}


\begin{abstract}
Quaternion algebra provides a natural representation of time-reversal-symmetric matrix structures in relativistic electronic-structure theory, whereas complementary time-reversal-antisymmetric structures extend this representation to complex quaternions, or biquaternions. Here, we explicitly exploit biquaternion algebra for fundamental objects, including operators, kinetically and magnetically balanced basis functions, and their expectation values, within a unified framework encompassing real, complex, and real-quaternion subalgebras as special cases. To realize this framework computationally, we have developed \textsc{HMATLIB}, a biquaternion matrix library implemented within the \textsc{ReSpect} package for pure CPU and hybrid CPU/GPU execution. A key development is a bilinear algorithm for matrix-valued biquaternion multiplication that reduces the number of real matrix--matrix multiplications from $64$ to $24$ compared with the conventional component-wise approach, while the biquaternion representation effectively doubles the maximum accessible matrix dimension under the same $32$-bit indexing constraint. Numerical benchmarks on modern CPU and GPU architectures demonstrate that the biquaternion formulation consistently outperforms its isomorphic complex-algebra counterpart. For the largest matrices reported, hybrid CPU/GPU execution accelerates bilinear matrix multiplication by approximately $5$--$8$ times over pure CPU execution and matrix diagonalization by approximately $12$ and $61$ times relative to CPU oneAPI MKL and NVHPC OpenBLAS, respectively.
These results establish biquaternion algebra as a general and computationally advantageous framework for modern relativistic electronic-structure calculations.
\end{abstract}

\maketitle


\section{Introduction}
\label{sec:introduction}

\noindent
The connection between time-reversal symmetry, Kramers' degeneracy, Kramers-restricted matrix structures, and quaternion algebra is well established in relativistic quantum chemistry and forms the basis of many modern two- and four-component electronic-structure implementations~\cite{Rosch1983,Saue1996,Dyall2007}. Quaternion representations have been employed, for instance, in Kramers-restricted 4-component relativistic Hartree--Fock~\cite{Saue1997} and Kohn--Sham~\cite{Saue2002,Kadek2019} theories, with a primary focus on matrix diagonalization~\cite{Rosch1983,Saue1997,Shiozaki2016}, the treatment of point-group symmetry~\cite{Saue1999}, or the localization of molecular orbitals~\cite{Senjean2021}. In these contexts, time-reversal-symmetric (TRS) matrices are naturally associated with closed-shell molecular systems and thus require only the algebra of \emph{real quaternions}, $\mathbb{H}_{\mathbb{R}}$.

What is less commonly emphasized is that time-reversal antisymmetric (TRA) matrices form a complementary space that can be identified with quaternions over the field of purely imaginary numbers, $i\mathbb{R}$. In this picture,
\begin{equation}
    \text{TRS matrices}
    \;\longleftrightarrow\;
    \mathbb{H}_{\mathbb{R}},
\end{equation}
whereas
\begin{equation}
    \text{TRA matrices}
    \;\longleftrightarrow\;
    i\mathbb{H}_{\mathbb{R}}.
\end{equation}
Combining the TRS and TRA components therefore leads naturally to the algebra of complex quaternions, or biquaternions~\cite{Lounesto2001},
\begin{equation}
    \mathbb{H}_{\mathbb{C}}
    =
    \mathbb{H}_{\mathbb{R}}
    \oplus
    i\mathbb{H}_{\mathbb{R}}.
\end{equation}
Biquaternions have a long-standing connection to relativistic physics, dating back to the work of Silberstein, who formulated classical electrodynamics and special relativity in terms of biquaternions~\cite{Silberstein1912,Silberstein1914}.

The relevance of biquaternions extends naturally to modern relativistic electronic-structure theory. As discussed in our previous work~\cite{ReSpect2020}, biquaternion structures arise not only in relativistic applications involving magnetic perturbations, such as Zeeman or hyperfine interactions~\cite{Repisky2010,Malkin2011,Komorovsky2013}, but also in applications requiring Kramers-unrestricted relativistic formulations, such as real-time electron dynamics~\cite{Repisky2015,Konecny2018} or noncollinear open-shell DFT and TDDFT~\cite{Komorovsky2019}. In all these cases, fundamental objects, such as charge-, spin-, and current-density matrices or restricted magnetic balance (RMB) basis functions~\cite{Komorovsky2008}, exhibit a biquaternion structure~\cite{ReSpect2020}. However, the corresponding matrix structures are commonly treated directly within complex algebra $\bbC$ through their characteristic block forms, without explicitly invoking their underlying biquaternion representation. The goal of this work, therefore, is to close this methodological gap by explicitly exploiting biquaternion algebra as a general, elegant, and computationally advantageous framework for relativistic electronic-structure calculations. Since the algebras of real numbers, complex numbers, and real quaternions are naturally embedded as subalgebras of the biquaternion algebra, this framework also provides a unified representation in simpler relativistic, scalar-relativistic, and nonrelativistic electronic-structure theory.

To realize this unified framework in practice and demonstrate its computational advantages over the algebra of complex numbers, we have developed \textsc{HMATLIB}, a matrix library based on biquaternion algebra, within our \textsc{ReSpect} DFT package~\cite{ReSpect2020,ReSpect2025}. Its core component is the \textsc{HMAT} module, which provides the algebraic operations over the field of biquaternions $\bbH_\bbC$ required in relativistic electronic-structure calculations. To overcome the high computational cost of straightforward component-wise multiplication of matrix-valued biquaternions, we have developed a bilinear multiplication algorithm that extends the original approach for real quaternions~\cite{quatpdt} to biquaternions and substantially reduces the computational time relative to complex-algebra multiplication. The \textsc{HMATLIB} supports both pure CPU and hybrid CPU/GPU parallel execution, with the latter backed by NVIDIA GPU-accelerated mathematical libraries, enabling efficient deployment on modern heterogeneous architectures.

The remainder of this article is organized as follows. Section~\ref{sec:theory} briefly introduces the algebra of biquaternions and establishes its connection to modern relativistic electronic-structure theory. Section~\ref{sec:implementation} describes the implementation of \textsc{HMATLIB}, including its technical aspects and the bilinear biquaternion matrix multiplication algorithm. Section~\ref{sec:results} presents numerical benchmarks demonstrating the computational advantages of biquaternion algebra over the conventional complex-algebra formulation. Finally, Section~\ref{sec:conclusion} summarizes the main findings and provides concluding remarks.

\section{Theory}
\label{sec:theory}

\subsection{Quaternions and Biquaternions}
\label{sec:theory/biquaternions}

\noindent
The quaternion algebra, denoted here as $\bbH_\bbR$, is a four-dimensional vector space over the real numbers $\bbR$ with a multiplicative structure. A quaternion number $q$ is generally defined as:
\begin{equation}
    q = a + b\qi + c\qj + d\qk \in \bbH_\bbR,
\end{equation}
where $a, b, c, d \in \bbR$, and $\qI, \qi, \qj, \qk$ are the quaternion basis elements satisfying
\begin{equation}
    \qi^2 = \qj^2 = \qk^2 = \qi\qj\qk = -1.
\end{equation}
Quaternions differ from real and complex numbers due to their non-commutative multiplication, meaning that for the product of two quaternions $p, q \in\bbH_\bbR$, we generally have $pq \neq qp$. Despite this non-commutativity, quaternion multiplication is associative  and a powerful mathematical structure for  relativistic quantum chemistry.

In a footnote to his seminal paper on magnetic electrons~\cite{jordan:quat}, Pauli credits Jordan with pointing out that the Pauli spin matrices
\begin{equation}
  \sigma_x =
  \begin{bmatrix}
  ~0~ & ~1~ \\
  ~1~ & ~0~ \\
  \end{bmatrix}
  \qquad
  \sigma_y =
  \begin{bmatrix}
  ~0~ & ~-i~ \\
  ~i~ & ~0~ \\
  \end{bmatrix}
  \qquad
  \sigma_z =
  \begin{bmatrix}
  ~1~ & ~0~ \\
  ~0~ & ~-1~ \\
  \end{bmatrix},
  \label{eq:PauliMatrices}
\end{equation}
multiplied by the negative imaginary unit,
can be identified with the quaternion units.
More precisely,
\begin{equation}
  -i\sigma_{x} \leftrightarrow \qi~~\qquad
  -i\sigma_{y} \leftrightarrow \qj~~\qquad
  -i\sigma_{z} \leftrightarrow \qk.
\end{equation}

An equally valid alternative mapping, better suited to the theory
discussed below, is obtained by considering a $2\times2$
complex matrix $M^{+}$ possessing the time-reversal symmetric (TRS)
structure
\begin{equation}
    M^+ =
    \begin{bmatrix}
        A & B \\
        -B^* & A^* \\
    \end{bmatrix}
    ,\qquad
    A, B\in\bbC .
\end{equation}
Expressing the matrix elements $A$ and $B$ in terms of their real ($R$) and imaginary ($I$) parts, the matrix can be expanded as
\begin{equation}
  M^+
  =
  I_{2} A_{R}
  + i\sigma_{z} A_{I}
  + i\sigma_{y} B_{R}
  + i\sigma_{x} B_{I}
  \in\bbC^{2\times2}
  \label{eq:TRS-in-C}
\end{equation}
which naturally suggests the mapping
\begin{equation}
  i\sigma_{z}\leftrightarrow\qi~~\qquad
  i\sigma_{y}\leftrightarrow\qj~~\qquad
  i\sigma_{x}\leftrightarrow\qk.
\label{eq:qmap1}
\end{equation}
The previous equation establishes the algebraic isomorphism
\begin{equation}
  \operatorname{span}_{\mathbb{R}}
  \{I_2,i\sigma_z,i\sigma_y,i\sigma_x\}
  \cong
  \bbH_{\bbR},
\end{equation}
allowing the complex Pauli-matrix representation of
Eq.~\eqref{eq:TRS-in-C} to be replaced by the equivalent, yet more
compact, quaternion representation
\begin{equation}
  M^+
  \cong
  A_{R}\qI + A_{I}\qi + B_{R}\qj + B_{I}\qk \in \bbH_{\bbR},
  \label{QMapM+}
\end{equation}
with $A_{R}, A_{I}, B_{R}, B_{I}\in\bbR$.

An analogous analysis can be carried out for time-reversal antisymmetric
(TRA) matrices of the general structure
\begin{equation}
    M^- =
    \begin{bmatrix}
        C & D \\
        D^* & -C^* \\
    \end{bmatrix}
    ,\qquad
    C, D\in\bbC.
\end{equation}
Their complex Pauli-matrix representation,
\begin{equation}
  M^-
  =
  i\big[
  I_{2} C_{I}
  - i\sigma_{z} C_{R}
  + i\sigma_{y} D_{I}
  - i\sigma_{x} D_{R}
  \big]
  \in\bbC^{2\times2},
  \label{eq:TRA-in-C}
\end{equation}
has expansion coefficients that are purely imaginary, i.e., elements of
$i\mathbb{R}$. This follows naturally from the observation that a TRA
matrix can be converted into a TRS matrix by extracting an imaginary phase~\cite{saue:linres}. Again, an equivalent but more compact
quaternion representation exists:
\begin{equation}
   M^-
   \cong
   i\big[
   C_{I} \qI + (-C_{R}) \qi + D_{I} \qj + (-D_{R}) \qk
   \big] \in i\bbH_{\bbR}.
   \label{QMapM-}
\end{equation}

What is less commonly emphasized is that any general $2\times2$ complex
matrix
\begin{equation}
    M =
    \begin{bmatrix}
      ~E~ & ~F~ \\
      ~G~ & ~H~ \\
    \end{bmatrix}
    ,\qquad
    E, F, G, H\in\bbC
\end{equation}
can be uniquely decomposed into TRS and TRA contributions:
\begin{align}
  M
  &=
  M^+ + M^-
  \nonumber
  \\
  &=
  I_{2}\bigl(A_{R}+iC_{I}\bigr)
  +i\sigma_{z}\bigl(A_{I}-iC_{R}\bigr)
  \nonumber
  \\
  &
  +i\sigma_{y}\bigl(B_{R}+iD_{I}\bigr)
  +i\sigma_{x}\bigl(B_{I}-iD_{R}\bigr)
  \in\bbC^{2\times2}.
  \label{eq:TRG-in-C}
\end{align}
This representation is isomorphic to the algebra of
quaternions over field of complex numbers ($\bbH_{\bbC}$), also known as the algebra of \textit{complex quaternions} or \textit{biquaternions},
\begin{equation}
  \operatorname{span}_{\bbC}
  \{I_2,i\sigma_z,i\sigma_y,i\sigma_x\}
  \cong
  \bbH_{\bbC},
\end{equation}
allowing the matrix $M$ in Eq.~\eqref{eq:TRG-in-C} to be represented as a
biquaternion,
\begin{equation}
\begin{aligned}
   M
   &\cong
   \bigl(A_{R}+iC_{I}\bigr)\qI
   +\bigl(A_{I}-iC_{R}\bigr)\qi
   \\[0.2cm]
   &\quad
   +\bigl(B_{R}+iD_{I}\bigr)\qj
   +\bigl(B_{I}-iD_{R}\bigr)\qk
   \in\bbH_{\bbC}.
\end{aligned}
\label{QMapM}
\end{equation}
The complex quantities $A$, $B$, $C$, and $D$ are obtained from the
matrix elements of $M$ as
\begin{equation}
  \begin{aligned}
  A &= \frac{E+H^*}{2},
    &\qquad
  B &= \frac{F-G^*}{2},
  \\[0.2cm]
  C &= \frac{E-H^*}{2},
    &\qquad
  D &= \frac{F+G^*}{2}.
  \end{aligned}
\end{equation}
We note that a quaternion algebra over a field $F$ can be either a division algebra or an algebra isomorphic to the algebra of all $2\times2$ matrices with entries in $F$, but not both~\cite{Lewis2006}. The real quaternion algebra is a division algebra and thus cannot be isomorphic to the algebra of all $2\times2$ real matrices, whereas the complex quaternion algebra is not a division algebra and is therefore isomorphic to the algebra of all $2\times2$ complex matrices.

At this point it is convenient to introduce a compact notation in which a quaternion is expressed as the sum of its scalar and vector parts,
\begin{equation}
  \label{eq:defQ}
  q
  = s + \boldsymbol{v}\qdot\qv
  = s + v_z\qi + v_y\qj + v_x\qk.
\end{equation}
where appears the \emph{reversed} quaternion vector
\begin{equation}
  \qv=(\qk,\qj,\qi),
\end{equation}
in order to comply with the mapping Eq.~\eqref{eq:qmap1}.
The product of two biquaternions, written in the scalar--vector form
$q_1=s_1+\boldsymbol{v}_1\qdot\qv\in\bbH_{\bbC}$ and
$q_2=s_2+\boldsymbol{v}_2\qdot\qv\in\bbH_{\bbC}$, where
$s_{1},s_{2}\in\mathbb{C}$ and
$\boldsymbol{v}_{1},\boldsymbol{v}_{2}\in\mathbb{C}^3$,
is given by
\begin{equation}
  q_1q_2
  =
  \left(s_1s_2-\boldsymbol{v}_1\cdot\boldsymbol{v}_2\right)
  +
  \left(
  s_1 \boldsymbol{v}_2
  +\boldsymbol{v}_1 s_2
  -\boldsymbol{v}_1\times\boldsymbol{v}_2
  \right)\qdot\qv\in\bbH_{\bbC}.
  \label{eq:QMultiplication}
\end{equation}
The noncommutative nature of quaternion multiplication is reflected in
the cross-product term, whose sign changes upon interchanging
$\boldsymbol{v}_1$ and $\boldsymbol{v}_2$.

All results derived above for biquaternions extend naturally to matrix-valued coefficients. We therefore define a matrix-valued biquaternion and its Hermitian adjoint (conjugate transpose) by
\begin{equation}
\begin{alignedat}{1}
\smat{Q}
&= \smat{S} + \boldsymbol{\smat{V}}\qdot\qv
 = \smat{S} + \smat{V}_z\qi + \smat{V}_y\qj + \smat{V}_x\qk
\in\bbH_{\bbC}^{n\times m}
\\
\smat{Q}^\dagger
&= \smat{S}^\dagger - \boldsymbol{\smat{V}}^\dagger\qdot\qv
 = \smat{S}^\dagger - \smat{V}_z^\dagger\qi - \smat{V}_y^\dagger\qj - \smat{V}_x^\dagger\qk
\in\bbH_{\bbC}^{m\times n}.
\end{alignedat}\label{eq:Q}
\end{equation}
where $\boldsymbol{\smat{V}}=(\smat{V}_x,\smat{V}_y,\smat{V}_z)$ and $\smat{S}, \smat{V}_x, \smat{V}_y, \smat{V}_z\in\mathbb{C}^{n\times m}$.
Here, $\bbH_{\bbC}^{n\times m}$ denotes all matrix-valued
biquaternions whose scalar and vector components belong to
$\mathbb{C}^{n\times m}$.
The product of two matrix-valued biquaternions
$\smat{Q}_1\in\bbH_{\bbC}^{n\times m}$ and
$\smat{Q}_2\in\bbH_{\bbC}^{m\times p}$
is given by
\begin{align}
  \smat{Q}_1\smat{Q}_2
  &=
  \left(
    \smat{S}_1\smat{S}_2
    -
    \boldsymbol{\smat{V}}_1\cdot\boldsymbol{\smat{V}}_2
  \right)
  \nonumber\\
  &+
  \left(
    \smat{S}_1\boldsymbol{\smat{V}}_2
    +
    \boldsymbol{\smat{V}}_1\smat{S}_2
    -
    \boldsymbol{\smat{V}}_1\times\boldsymbol{\smat{V}}_2
  \right)\qdot\qv
  \in
  \bbH_{\bbC}^{n\times p}.
  \label{eq:MatrixQMultiplication}
\end{align}
The multiplication rule has exactly the same formal structure as
Eq.~\eqref{eq:QMultiplication}; only the scalar and vector coefficients
have been promoted to matrices and thus they do not commute. The dot and cross products preserve the
order of matrix multiplication and are understood componentwise.

Under the established isomorphism, a matrix-valued biquaternion $Q$, Eq.~\eqref{eq:Q}, is equivalent to the Pauli-matrix representation
\begin{equation}\label{eq:Q_Pauli}
Q\cong M=I_2\otimes S+ i\boldsymbol{\sigma}\otimes\boldsymbol{V},
\end{equation}
where we use the Kronecker product ($\otimes$) with the notation
\begin{equation}
  \boldsymbol{\sigma}\otimes\boldsymbol{V}
  \equiv
  \sigma_x\otimes V_x
  +\sigma_y\otimes V_y
  +\sigma_z\otimes V_z.
\end{equation}
Note that the Kronecker product of two matrices $A$ and $B$ is noncommutative. The two orderings are related by permutation matrices $P$ and $Q$,
\begin{equation}\label{eq:Kronecker_order}
A \otimes B = P^T(B \otimes A)Q.
\end{equation}
An example will be seen in the next section.

We now calculate
\begin{align}
  M_1M_2
  &=
  \left(
    I_2\otimes S_1
    +i\boldsymbol{\sigma}\otimes\boldsymbol{V}_1
  \right)
  \left(
    I_2\otimes S_2
    +i\boldsymbol{\sigma}\otimes\boldsymbol{V}_2
  \right)
  \nonumber\\
  &=
  I_2\otimes
  \left(
    S_1S_2-\boldsymbol{V}_1\cdot\boldsymbol{V}_2
  \right)
  \nonumber\\
  &+
  i\boldsymbol{\sigma}\otimes
  \left(
    S_{1}\boldsymbol{V}_{2}
    +\boldsymbol{V}_{1}S_2
    -\boldsymbol{V}_1\times\boldsymbol{V}_2
  \right),
\end{align}
where we used the Kronecker-product identity
\begin{equation}
  (A\otimes B)(C\otimes D)=AC\otimes BD
\end{equation}
as well as the Dirac identity~\cite[Eq.~(16)]{dirac:eq1}
\begin{equation}
  (\boldsymbol{\sigma}\cdot\boldsymbol{a})
  (\boldsymbol{\sigma}\cdot\boldsymbol{b})
  =
  (\boldsymbol{a}\cdot\boldsymbol{b})I_2
  +
  i\boldsymbol{\sigma}\cdot
  (\boldsymbol{a}\times\boldsymbol{b}).
\end{equation}
The result indeed corresponds to Eq.~\eqref{eq:MatrixQMultiplication}, confirming the equivalence $M_1M_2\cong \smat{Q}_1\smat{Q}_2$ between the Pauli-matrix representation and the more compact quaternion one.

To conclude, the algebra of biquaternions provides a unified, compact, and computationally advantageous framework for representing arbitrary matrices arising in both relativistic and nonrelativistic electronic-structure theory, as the algebras of real numbers, complex numbers, and quaternions can all be embedded as subalgebras of the biquaternion algebra $\bbH_{\bbC}$. This observation forms the mathematical foundation for the development of a universal mathematical library, \textsc{HMATLIB}. The connection between biquaternion algebra and relativistic quantum chemistry is demonstrated explicitly in the following section.
\subsection{Quaternions and Relativistic Quantum Chemistry}
\label{sec:theory/HRinQCH}

\noindent
The starting point for our discussion is the one-electron Dirac Hamiltonian (in SI units):
\begin{equation}
  \label{eq:Hamiltonian-4c-1e}
h^\text{D} = \beta' m c^2 + c (\vec{\alpha} \cdot \vec{p})  + V
\end{equation}
Here, $c$ denotes the speed of light, $\vec{p}$ is the linear momentum operator, and $V$ represents the interaction with the electrostatic scalar potential due to the atomic nuclei. The Dirac matrices $\vec{\alpha}$ and $\beta'$ are assumed to be given in the representation
\begin{equation}
  \alpha_k =
  \begin{bmatrix}
      ~0_2~      & ~\sigma_k~ \\
      ~\sigma_k~ & ~0_2~ \\
  \end{bmatrix}
  \qquad
  \beta' =
  \begin{bmatrix}
      ~0_{2}~ &  ~0_2~ \\
      ~0_2~   & ~-2\text{I}_2~ \\
  \end{bmatrix}
  \qquad
  k \in x, y, z
\end{equation}
where $0_2$ and $\text{I}_2$ denote the $2\times2$ zero and identity matrices, respectively, and $\sigma_k$ are the $2\times2$ Pauli spin matrices defined in Eq.~\eqref{eq:PauliMatrices}.
The TRS structure of this $4\times 4$ Hamiltonian can be brought out by resorting~\cite{Saue1997}
\begin{equation}
  \left[\begin{array}{c}
    \psi^L\\
    \psi^S
  \end{array}\right]
  =
  \left[\begin{array}{c}
    \psi^{L\alpha}\\
    \psi^{L\beta}\\
    \psi^{S\alpha}\\
    \psi^{S\beta}
  \end{array}\right]
  \quad\rightarrow\quad
  \left[\begin{array}{c}
    \psi^{\alpha}\\
    \psi^{\beta}
  \end{array}\right]
  =
  \left[\begin{array}{c}
    \psi^{L\alpha}\\
    \psi^{S\alpha}\\
    \psi^{L\beta}\\
    \psi^{S\beta}
  \end{array}\right]
  .
  \label{eq:LS2ab}
\end{equation}
Mathematically, this corresponds to the transformation of the Dirac Hamiltonian by means of an orthogonal permutation matrix $P$, which satisfies $P^{T} = P^{-1}$,
\begin{equation}
    h^\text{D} \rightarrow P^{T} h^\text{D} P,
    \qquad
    P =
    \begin{bmatrix}
        ~~1~~ & ~~0~~ & ~~0~~ & ~~0~~ \\[-0.1cm]
         ~0   &  ~0~  &  ~1~  &  ~0~ \\[-0.1cm]
         ~0   &  ~1~  &  ~0~  &  ~0~ \\[-0.1cm]
         ~0   &  ~0~  &  ~0~  &  ~1~ \\
    \end{bmatrix}
    \label{permutation-Dirac}
\end{equation}
and results in the transformed Dirac Hamiltonian with the time-reversal symmetric structure
\begin{equation}
    P^{T} h^\text{D} P
    =
    \begin{bmatrix}
         A   & B \\
        -B^* & A^* \\
    \end{bmatrix},
    \label{trs-matrix}
\end{equation}
where its elements read
\begin{equation}
\begin{aligned}
    A
    & =
    \begin{bmatrix}
      V       & -ic\hbar\nabla_{\!z} \\
      -ic\hbar\nabla_{\!z} & V-2mc^2 \\
    \end{bmatrix}
    \\[0.2cm]
    B
    & =
    \begin{bmatrix}
      0 & -ic\hbar\nabla_{\!x}  -  c\hbar\nabla_{\!y} \\
      -ic \hbar\nabla_{\!x}  -  c\hbar\nabla_{\!y} & 0 \\
    \end{bmatrix}
    \label{a-and-b-in-trs-dirac-h}
\end{aligned}
\end{equation}
With the help of Eqs.~\eqref{QMapM+} and~\eqref{eq:defQ}, the transformed Dirac Hamiltonian can be written in the quaternion form as:
\begin{equation}
\begin{aligned}
    {^{\text{q}\!}}h^{\text{D}}
    & =
    \begin{bmatrix}
    V & 0\\
0 & V-2mc^{2}
    \end{bmatrix} +
    \begin{bmatrix}
    \boldsymbol{0} & -c\hbar\boldsymbol{\nabla}\\
    -c\hbar\boldsymbol{\nabla} & \boldsymbol{0}
    \end{bmatrix}
    \qdot\qv
    \label{dirac-h-in-q-form}
\end{aligned}
\end{equation}
The quaternion form of the Dirac Hamiltonian has an interesting structure: the scalar potential appears in the scalar part (the component associated with $\qI$) while the kinetic energy components are distributed across the vector part (the components associated with $\qi,\qj,\qk$). The equivalence of quaternion units mirrors the symmetry of the coordinate axes $(x, y, z)$. This equivalence also ensures that the quaternion Dirac Hamiltonian does not refer to any particular spin quantization axis, contrary to  the parent Dirac Hamiltonian.

From the discussion in the previous section it is clear that the Dirac Hamiltonian can also be written in the Pauli-matrix representation, Eq.~\eqref{eq:Q_Pauli}, after the reordering corresponding to Eq.~\eqref{eq:LS2ab}. The original form, Eq.~\eqref{eq:Hamiltonian-4c-1e}, is obtained by swapping the order of matrices in the Kronecker products. This is accordingly a special instance of Eq.~\eqref{eq:Kronecker_order}, with $Q=P$.

Similar to the Dirac Hamiltonian, one must also transform the 4c basis functions $X^{\text{RKB}}$ using the orthogonal permutation matrix $P$:
\begin{equation}
\begin{aligned}
   X_{\mu}^{\text{RKB}}(\vec{r})
   & \rightarrow
   P^{T}X_{\mu}^{\text{RKB}}(\vec{r})
   \in\bbC^{4\times4}(\bbR^{3})
   \\[0.2cm]
   X_{\mu}^{\text{RKB}}(\vec{r})
   & =
   \begin{bmatrix}
      \text{I}_{2} & 0_{2}
      \\[0.2cm]
      0_{2} & \frac{1}{2mc}(\vec{\sigma}\cdot\vec{p})
   \end{bmatrix}
   f_{\mu}(\vec{r})
   \in\bbC^{4\times4}(\bbR^{3})
\end{aligned}
\end{equation}
Here, $f_{\mu}(\vec{r})$ refers to a normalized scalar basis function with identical normalization for both the large ($\text{L}$) and small ($\text{S}$) basis components, respectively. The latter are related through the restricted kinetic balance (RKB) condition~\cite{Stanton1984,dyall:rbal}. After the transformation, the RKB basis also acquires the time-reversal symmetric structure~\cite{ReSpect2020}:
\begin{equation}
    P^{T}X_{\mu}^{\text{RKB}}(\vec{r})
    =
    \begin{bmatrix}
        A_{\mu}(\vec{r}) & B_{\mu}(\vec{r})
        \\[0.2cm]
        -B_{\mu}^*(\vec{r}) & A_{\mu}^*(\vec{r})
    \end{bmatrix}
\end{equation}
with the $2\times2$ complex functions $A_{\mu}(\vec{r})$ and $B_{\mu}(\vec{r})$ given by
\begin{equation}
  \begin{aligned}
    A_{\mu}(\vec{r})
    & =
    \begin{bmatrix}
        f_{\mu}(\vec{r}) & 0 \\[0.2cm]
        0 & -\frac{i\hbar}{2mc}\nabla_{\!z} f_{\mu}(\vec{r})
    \end{bmatrix}
    \in \bbC^{2 \times 2} (\bbR^3)
    \\[0.2cm]
    B_{\mu}(\vec{r})
    & =
    \begin{bmatrix}
        0 & 0 \\[0.2cm]
        0 & -\frac{\hbar}{2mc}(\nabla_{\!y} + i \nabla_{\!x}) f_{\mu}(\vec{r})
    \end{bmatrix}
    \in \bbC^{2 \times 2} (\bbR^3)
    \label{a-and-b-in-trs-basis}
  \end{aligned}
\end{equation}
\hspace{0.2cm}
With the help of equations~\eqref{QMapM+} and~\eqref{eq:defQ}, the transformed 4c RKB basis functions can be written in the quaternion form as:
\begin{equation}
  |{^{\text{q}\!}}X_{\mu}^{\text{RKB}}\rangle
  =
  \begin{bmatrix}
  |f_{\mu}\rangle & 0\\
  0 & 0
  \end{bmatrix}
  +
  \begin{bmatrix}
  \boldsymbol{0} & \boldsymbol{0}\\
  \boldsymbol{0} & -\frac{\hbar}{2mc}\boldsymbol{\nabla}|f_{\mu}\rangle
  \end{bmatrix}
  \qdot\qv
  \in \bbH_{\bbR}^{2\times 2}(\bbR^3)
\end{equation}
Throughout this section, we shall use the Dirac bra-ket formalism, which notably allows us to employ the turnover rule in the form\cite[p.26]{Dirac:1958:PQM}
\begin{equation}
  \langle Q\psi|=\langle \psi|Q^{\dagger}.
\end{equation}

\noindent
Magnetic fields are included through minimal substitution~\cite{Gell1956}
\begin{equation}
  \boldsymbol{p}\rightarrow\boldsymbol{p}+e\boldsymbol{A};\quad\Rightarrow\quad\hbar\boldsymbol{\nabla}\rightarrow\hbar\boldsymbol{D}=\hbar\boldsymbol{\nabla}+ie\boldsymbol{A},
\end{equation}
where $\boldsymbol{D}$ is the covariant derivative on flat space-time.
We then obtain
\begin{equation}
  {^{\text{q}\!}}h^{\text{D}}_{\left(\varphi,\boldsymbol{A}\right)}
  =
  \left[\begin{array}{cc}
    V & 0\\
    0 & V-2mc^{2}
  \end{array}
  \right]
  +\left[\begin{array}{cc}
     \boldsymbol{0} & -c\hbar\boldsymbol{D}\\
    -c\hbar\boldsymbol{D} & \boldsymbol{0}
\end{array}\right]\qdot\qv.
\label{rkb-in-q-form}
\end{equation}
With the introduction of magnetic fields, we employ restricted magnetic
balance (RMB)~\cite{saue:dso,Komorovsky2008,Repisky2009,Komorovsky2010}
\begin{equation}
  |{^{\text{q}\!}}X_{\mu}^{\text{RMB}}\rangle
  =
  \begin{bmatrix}
    |f_{\mu}\rangle & 0\\
    0 & 0
  \end{bmatrix}
  +
  \begin{bmatrix}
    \boldsymbol{0} & \boldsymbol{0}\\
    \boldsymbol{0} & -\frac{\hbar}{2mc}\boldsymbol{D}|f_{\mu}\rangle
  \end{bmatrix}\qdot\qv.
\end{equation}
We note that
\begin{equation}
   \boldsymbol{D}^{\dagger}=-\boldsymbol{D},
\end{equation}
hence the adjoint is
\begin{equation}
  \langle {^{\text{q}\!}}X_{\mu}^{\text{RMB}}|
  =
  \begin{bmatrix}
    \langle f_{\mu}| & 0\\
    0 & 0
  \end{bmatrix}
  -
  \begin{bmatrix}
     \boldsymbol{0} & \boldsymbol{0}\\
     \boldsymbol{0} & \frac{\hbar}{2mc}\langle f_{\mu}|\boldsymbol{D}
  \end{bmatrix}
  \qdot\qv
\end{equation}
We now calculate
\begin{equation}
\begin{aligned}
  {^{\text{q}\!}}h^{\text{D}}_{(\varphi,\boldsymbol{A})}
  |{^{\text{q}\!}}X_{\nu}^{\text{RMB}}\rangle
  &=
  \begin{bmatrix}
    V|f_{\nu}\rangle & -\frac{\hbar^2}{2m}(\boldsymbol{D}\cdot\boldsymbol{D})|f_{\nu}\rangle
    \\[0.2cm]
    0 & 0
  \end{bmatrix}
  \\[0.2cm]
  &+
  \begin{bmatrix}
    \boldsymbol{0}
    &
    -\frac{\hbar^{2}}{2mc}\left(\boldsymbol{D}\times\boldsymbol{D}\right)|f_{\nu}\rangle
    \\[0.2cm]
    -c\hbar\boldsymbol{D}|f_{\nu}\rangle
    &
    -\frac{\hbar}{2mc}\left(V-2mc^{2}\right)\boldsymbol{D}|f_{\nu}\rangle
  \end{bmatrix}
  \qdot\qv.
\end{aligned}
\end{equation}
and next
\begin{widetext}
\begin{equation}
\begin{aligned}
  \langle {^{\text{q}\!}}X_{\mu}^{\text{RMB}}|
  {^{\text{q}\!}}h^\text{D}_{(\varphi,\boldsymbol{A})}|
  {^{\text{q}\!}}X_{\nu}^{\text{RMB}}\rangle
  & =
  \begin{bmatrix}
    \langle f_{\mu}|V|f_{\nu}\rangle
    &
    -\frac{\hbar^{2}}{2m}\langle f_{\mu}|\boldsymbol{D}^{2}|f_{\nu}\rangle
    \\[0.2cm]
    -\frac{\hbar^2}{2m}\langle f_{\mu}|\boldsymbol{D}^{2}|f_{\nu}\rangle
    &
    -\frac{\hbar^{2}}{4m^{2}c^{2}}\langle f_{\mu}|\boldsymbol{D}\cdot\left(V-2mc^{2}\right)\boldsymbol{D}|f_{\nu}\rangle
  \end{bmatrix}
  \\[0.2cm]
  & +
  \begin{bmatrix}
    \boldsymbol{0}
    &
    -\frac{\hbar^{2}}{2mc}\langle f_{\mu}|\left(\boldsymbol{D}\times\boldsymbol{D}\right)|f_{\nu}\rangle
    \\[0.2cm]
    -\frac{\hbar^{2}}{2m}\langle f_{\mu}|\boldsymbol{D}\times\boldsymbol{D}|f_{\nu}\rangle &
    -\frac{\hbar^{2}}{4m^{2}c^{2}}\langle f_{\mu}|\boldsymbol{D}\times\left(V-2mc^{2}\right)\boldsymbol{D}|f_{\nu}\rangle
  \end{bmatrix}
  \qdot\qv.
\end{aligned}
\end{equation}
\end{widetext}
We also need metric elements
\begin{equation}
\begin{aligned}
  \langle {^{\text{q}\!}}X_{\mu}^{\text{RMB}}|
  {^{\text{q}\!}}X_{\nu}^{\text{RMB}}\rangle
  &=
  \begin{bmatrix}
    \langle f_{\mu}|f_{\nu}\rangle & 0
    \\[0.2cm]
    0 & -\frac{\hbar^{2}}{4m^{2}c^{2}}\langle f_{\mu}|\boldsymbol{D}^{2}|f_{\nu}\rangle
  \end{bmatrix}
  +
  \begin{bmatrix}
    \boldsymbol{0} & \boldsymbol{0}
    \\[0.2cm]
    \boldsymbol{0} & -\frac{\hbar^{2}}{4m^{2}c^{2}}\langle f_{\mu}|\boldsymbol{D}\times\boldsymbol{D}|f_{\nu}\rangle
  \end{bmatrix}
  \qdot\qv.
\end{aligned}
\end{equation}
Further simplifications are obtained using
\[
\hbar^{2}\left(\boldsymbol{D}\times\boldsymbol{D}\right)=ie\hbar\boldsymbol{B}.
\]
 We also note
\[
\hbar^{2}\boldsymbol{D}^{2}=\hbar^{2}\nabla^{2}+ie\hbar\left(\boldsymbol{\nabla}\cdot\boldsymbol{A}+\boldsymbol{A}\cdot\boldsymbol{\nabla}\right)-e^{2}A^{2},
\]
which for Coulomb gauge ($\boldsymbol{\nabla}\cdot\boldsymbol{A}=0$)
reduces to
\[
\hbar^{2}\boldsymbol{D}^{2}=\hbar^{2}\nabla^{2}+2ie\hbar\boldsymbol{A}\cdot\boldsymbol{\nabla}-e^{2}A^{2}.
\]
For $\boldsymbol{A}=\boldsymbol{0}$ the matrix elements reduce to
\begin{equation}
  \begin{aligned}
  \langle {^{\text{q}\!}}X_{\mu}^{\text{RKB}}|
  {^{\text{q}\!}}h^{\text{D}}_{(\varphi,\boldsymbol{0})}|
  {^{\text{q}\!}}X_{\nu}^{\text{RKB}}\rangle
  & =
  \begin{bmatrix}
  \langle f_{\mu}|V|f_{\nu}\rangle
  & -\frac{\hbar^{2}}{2m}\langle f_{\mu}|\nabla^{2}|f_{\nu}\rangle
  \\[0.2cm]
  -\frac{\hbar^{2}}{2m}\langle f_{\mu}|\nabla^{2}|f_{\nu}\rangle
  &
  -\frac{\hbar^{2}}{4m^{2}c^{2}}\langle f_{\mu}|\boldsymbol{\nabla}\cdot\left(V-2mc^{2}\right)\boldsymbol{\nabla}|f_{\nu}\rangle
  \end{bmatrix}
  \\
  & +
  \begin{bmatrix}
    \boldsymbol{0} & \boldsymbol{0}
    \\[0.2cm]
    \boldsymbol{0} & -\frac{\hbar{^{2}}}{4m^{2}c^{2}}\langle f_{\mu}|\boldsymbol{\nabla}\times V\boldsymbol{\nabla}|f_{\nu}\rangle
  \end{bmatrix}
  \qdot\qv
  \end{aligned}
\end{equation}
and
\begin{equation}
  \langle {^{\text{q}\!}}X_{\mu}^{\text{RKB}} | {^{\text{q}\!}}X_{\nu}^{\text{RKB}}\rangle
  =
  \begin{bmatrix}
     \langle f_{\mu}|f_{\nu}\rangle & 0\\
     0 & -\frac{\hbar^{2}}{4m^{2}c^{2}}\langle f_{\mu}|\nabla^{2}|f_{\nu}\rangle
  \end{bmatrix}
  .
\end{equation}
A selected list of quantum chemical operators, together with their inner products, is presented in the biquaternion algebra formalism in Tables~\ref{tab:biquat-table-operators} and~\ref{tab:biquat-table-inner-pdts}.
%
%
\begingroup
\small
\setlength{\tabcolsep}{3pt}
\renewcommand{\arraystretch}{1.3}

\begin{longtable}{cll}

\caption{\label{tab:biquat-table-operators}
Identification of zero and nonzero biquaternion components of selected quantum chemical operators.}
\\

\hline\hline
\textbf{Biquaternion~~}
& \textbf{Name/Description}
& \textbf{Operator} \\
\hline
\endfirsthead

\multicolumn{3}{c}{%
\tablename~\thetable\ (continued)}\\
\hline\hline
\textbf{Biquaternion~~}
& \textbf{Name/Description}
& \textbf{Operator} \\
\hline
\endhead

\hline
\multicolumn{3}{r}{Continued on next page}\\
\endfoot

\hline\hline
\endlastfoot

[1,0,0,0,0,0,0,0]
& Four-component identity operator
& $\hat{1}$ \\

[1,0,0,0,0,0,0,0]
& Scalar potential
& $\hat{V}$ \\

[1,0,0,0,0,0,0,0]
& Real scalar function
& $f_\mu(\vec{r})$ \\

[0,0,0,0,1,0,0,0]
& Non-relativistic momentum operator
& $\vec{p}$ \\

[0,0,0,0,1,0,0,0]
& Angular momentum operator
& $\vec{L}$ \\

[1,0,0,0,1,0,0,0]
& London atomic orbitals
& $f^\text{LAO}_\mu(\vec{r})$ \\

[0,1,1,1,0,0,0,0]
& One-electron spin--orbit operator
& $\xi(\vec{r}) \vec{L}\cdot\vec{\sigma}$ \\

[1,1,1,1,0,0,0,0]
& Dirac operator in the absence of magnetic fields
& $\hat{h}^\text{D}$ \\

[1,1,1,1,0,0,0,0]
& Restricted kinetically balanced basis
& $X_{\mu}^{\text{RKB}}(\vec{r})$ \\

[1,0,0,0,1,1,1,1]
& Gordon-decomposed current density operator
& $-\frac{1}{2}(\vec{p}-\cancel{\vec{p}})\beta
   -\frac{1}{2}\vec{\nabla}\times\vec{\Sigma}\beta
   -\frac{1}{c}\vec{A}\beta$ \\

[0,0,0,0,0,1,1,1]
& Four-component current density operator
& $-c\vec{\alpha}$ \\

[0,0,0,0,0,1,1,1]
& Four-component magnetic-field interaction Hamiltonian
& $\vec{\alpha}\cdot\vec{A}$ \\

[0,0,0,0,0,1,1,1]
& Vector of Pauli matrices
& $\vec{\sigma}$ \\

[0,0,0,0,0,1,1,1]
& Magnetic component of the RMB basis
& $X_{\mu}^{\text{RMB},A}(\vec{r})$ \\

[0,0,0,0,1,1,1,1]
& Four-component total angular momentum
& $\vec{L} \hat{1} + \frac{1}{2}\vec{\Sigma}$ \\

[0,0,0,0,1,1,1,1]
& Magnetic-field derivative of the RMB--GIAO basis
& $X_{\mu}^{\text{RMB--GIAO},B_u}(\vec{r})$ \\

[1,1,1,1,0,1,1,1]
& Restricted magnetically balanced basis
& $X_{\mu}^{\text{RMB}}(\vec{r})$ \\

[1,1,1,1,1,1,1,1]
& Combined London and RMB basis
& $X_{\mu}^{\text{RMB--GIAO}}(\vec{r})$ \\

\end{longtable}
\endgroup

%
%
\begin{table*}
\caption{\label{tab:biquat-table-inner-pdts}
Identification of zero and nonzero biquaternion components of selected quantum chemical operator matrices.
}
\begin{ruledtabular}
\begin{tabular}{ccc}
\multicolumn{2}{c}{\textbf{Biquaternion components}} & \textbf{Operator} \\
\hline
&\\[-5pt]

$\langle X_{\mu}^{\text{RKB}} | \hat{O} | X_{\nu}^{\text{RKB}} \rangle$ &
$\langle X_{\mu}^{\text{RKB}} | \hat{O} | X_{\nu}^{\text{RMB},A} \rangle$ + h.c. &
$\hat{O}$ \\
\hline

[1,0,0,0,0,0,0,0] & [0,0,0,0,1,1,1,1] & $\hat{1}$ \\

[1,1,1,1,0,0,0,0] & [0,0,0,0,1,1,1,1] & $\hat{h}^\text{D}$ \\

[0,0,0,0,1,1,1,1] & [1,1,1,1,0,0,0,0] & $-c\vec{\alpha}$ \\

[0,0,0,0,1,1,1,1] & [1,0,0,0,0,0,0,0] & $\vec{\alpha}\cdot\vec{A}$ \\

[0,0,0,0,0,1,1,1] & [1,1,1,1,0,0,0,0] & $\vec{\Sigma}$ \\

$ (X_{\mu}^{\text{RKB}})^\dagger \hat{O} X_{\nu}^{\text{RKB}}$ &
$ (X_{\mu}^{\text{RKB}})^\dagger \hat{O} X_{\nu}^{\text{RMB},A}$ + h.c. &
$\hat{O}$ \\
\hline

[1,1,1,1,0,0,0,0] & [0,0,0,0,1,1,1,1] & $\hat{1}$ \\

[0,0,0,0,1,1,1,1] & [1,1,1,1,0,0,0,0] & $-c\vec{\alpha}$ \\

[0,0,0,0,1,1,1,1] & [1,1,1,1,0,0,0,0] & $\vec{\Sigma}$ \\

\end{tabular}
\end{ruledtabular}
\end{table*}

\section{Implementation Details}
\label{sec:implementation}
\noindent
The mathematical framework introduced above forms the foundation of \textsc{HMATLIB}, a matrix library developed within our \textsc{ReSpect} DFT package~\cite{ReSpect2020,ReSpect2025}. Its core component is the \textsc{HMAT} module, which provides the algebraic operations over the field of biquaternions ($\bbH_{\bbC}$) required in relativistic electronic-structure calculations. Since the biquaternion algebra naturally encompasses real, complex, and quaternion subalgebras, the module is also well suited for simpler scalar-relativistic and nonrelativistic calculations. The module supports both pure CPU and hybrid CPU/GPU parallel execution, with the latter backed by NVIDIA GPU-accelerated mathematical libraries. Two additional modules, the real-algebra-based \textsc{RMAT} and complex-algebra-based \textsc{CMAT}, are included in the library and serve as reference implementations for the validation and performance benchmarking of \textsc{HMAT}.

Within \textsc{HMAT}, a general matrix-valued biquaternion $\qMb{q}{}\in\bbH_{\bbC}^{m\times n}$ is represented by eight real $m\times n$ matrices
\begin{align}
\qMb{q}{}
    & \equiv
    [~\qMb{}{1}, \;
      \qMb{}{2}, \;
      \qMb{}{3}, \;
      \qMb{}{4}, \;
      \qMb{}{5}, \;
      \qMb{}{6}, \;
      \qMb{}{7}, \;
      \qMb{}{8}] \nonumber
\\
    & =
    [~\mat{E}_\text{R}, \;
      \mat{E}_\text{I}, \;
      \mat{F}_\text{R}, \;
      \mat{F}_\text{I}, \;
      \mat{G}_\text{I}, \;
     -\mat{G}_\text{R}, \;
      \mat{H}_\text{I}, \;
     -\mat{H}_\text{R} ].
     \label{biquat-repres}
\end{align}
The components of $\qMb{q}{}$ represent the real and imaginary parts, denoted by the subscripts R and I, respectively, of four complex matrices $\mat{E}, \mat{F}, \mat{G}, \mat{H}\in\bbC^{m\times n}$ associated with the biquaternion basis elements $\qI$, $\qi$, $\qj$, and $\qk$:
\begin{equation}
\begin{aligned}
  \qMb{q}{}
  & =
  (\mat{E}_\text{R} + i\mat{G}_\text{I})\qI + (\mat{E}_\text{I} - i\mat{G}_\text{R})\qi
  \\
  &+
  (\mat{F}_\text{R} + i\mat{H}_\text{I})\qj + (\mat{F}_\text{I} - i\mat{H}_\text{R})\qk.
  \label{main-biquat-eqn}
\end{aligned}
\end{equation}
The isomorphism between the biquaternion and complex matrix algebras established in Section~II\,A relates the matrices $\mat{E}$ and $\mat{F}$ to the components of the complex time-reversal symmetric matrix $\mat{M}^{+}$,
\begin{equation}
\mat{M}^+ =
\begin{bmatrix}
    \mat{E} & \mat{F}
    \\[0.15cm]
    -\mat{F}^* & \mat{E}^*
\end{bmatrix}
\in\bbC^{2m\times2n},
\end{equation}
where $^*$ denotes complex conjugation. Similarly, the matrices $\mat{G}$ and $\mat{H}$ are related to the components of the complex time-reversal antisymmetric matrix $\mat{M}^{-}$,
\begin{equation}
\mat{M}^- =
\begin{bmatrix}
    \mat{G} & \mat{H}
    \\[0.15cm]
    \mat{H}^* & -\mat{G}^*
\end{bmatrix}
\in\bbC^{2m\times2n}.
\end{equation}
Note that the complex matrices $\mat{M}^{+}$ and $\mat{M}^{-}$ are twice as large as their defining matrices $\mat{E}, \mat{F}, \mat{G}, \mat{H}$. Since a general complex matrix $\mat{M}$ of size $2m\times2n$
\begin{equation}
\mat{M} =
\begin{bmatrix}
    \mat{A} & \mat{B}
    \\[0.15cm]
    \mat{C} & \mat{D}
\end{bmatrix},
\qquad
\mat{A}, \mat{B}, \mat{C}, \mat{D}\in\bbC^{m\times n}
\end{equation}
can always be decomposed into its time-reversal symmetric and time-reversal antisymmetric parts,
\begin{equation}
\mat{M} = \mat{M}^+ + \mat{M}^-,
\end{equation}
one obtains the following explicit relations defining the isomorphic mapping used within our \textsc{HMAT} module:
\begin{equation}
\begin{aligned}
  \mat{E} &= \frac{\mat{A} + \mat{D}^*}{2} \qquad
  \mat{F}  = \frac{\mat{B} - \mat{C}^*}{2}
  \\
  \mat{G} &= \frac{\mat{A} - \mat{D}^*}{2} \qquad
  \mat{H}  = \frac{\mat{B} + \mat{C}^*}{2}
  \label{eq:mapping-formulas}
\end{aligned}
\end{equation}

The \textsc{HMAT} module offers a basic set of mathematical operations, namely scaling by a complex factor $\alpha\in\bbC$:
\begin{equation}
\begin{aligned}
\alpha\qMb{q}{}
=
\big[ &
\alpha_{\text{R}}\qMb{}{1} - \alpha_{\text{I}}\qMb{}{5},\,
\alpha_{\text{R}}\qMb{}{2} - \alpha_{\text{I}}\qMb{}{6},
\\ &
\alpha_{\text{R}}\qMb{}{3} - \alpha_{\text{I}}\qMb{}{7},\,
\alpha_{\text{R}}\qMb{}{4} - \alpha_{\text{I}}\qMb{}{8},
\\ &
\alpha_{\text{R}}\qMb{}{5} + \alpha_{\text{I}}\qMb{}{1},\,
\alpha_{\text{R}}\qMb{}{6} + \alpha_{\text{I}}\qMb{}{2},
\\ &
\alpha_{\text{R}}\qMb{}{7} + \alpha_{\text{I}}\qMb{}{3},\,
\alpha_{\text{R}}\qMb{}{8} + \alpha_{\text{I}}\qMb{}{4} \big],
\end{aligned}
\end{equation}
addition of two biquaternions $\qMb{q}{},\qNb{q}{}\in\bbH_{\bbC}^{m\times n}$:
\begin{equation}
    \begin{aligned}
    \qMb{q}{}   + \qNb{q}{}
    =
    [\qMb{}{1} &+ \qNb{}{1},
     \qMb{}{2}  + \qNb{}{2},
     \qMb{}{3}  + \qNb{}{3},
     \qMb{}{4}  + \qNb{}{4},
     \\
     \qMb{}{5} &+ \qNb{}{5},
     \qMb{}{6}  + \qNb{}{6},
     \qMb{}{7}  + \qNb{}{7},
     \qMb{}{8}  + \qNb{}{8}],
    \end{aligned}
\end{equation}
biquaternion trace for square $\qMb{q}{}\in\bbH_{\bbC}^{m\times m}$:
\begin{equation}
    \text{Tr}[\qMb{q}{}]
    =
    2\big(\text{Tr}[\qMb{}{1}] + \, i\,\text{Tr}[\qMb{}{5}]\big),
\end{equation}
as well as multiplication and diagonalization. With the exception of diagonalization, all of these operations seamlessly support the biquaternion transposition ($T$), complex conjugation ($*$) and Hermitian conjugation ($\dag$). These transformations require only a simple matrix transposition and a sign change of individual components with respect to the original layout defined in Eq.~(\ref{biquat-repres}), namely,
transposition:
    \begin{equation}
        \qMb{q}{}^{T} =
       [ \qMb{}{1}^{T}, \;
         \qMb{}{2}^{T}, \;
        -\qMb{}{3}^{T}, \;
         \qMb{}{4}^{T}, \;
         \qMb{}{5}^{T}, \;
         \qMb{}{6}^{T}, \;
        -\qMb{}{7}^{T}, \;
         \qMb{}{8}^{T}  ]
    \end{equation}
complex conjugation:
    \begin{equation}
        \qMb{q}{}^{*} =
       [ \qMb{}{1}, \;
        -\qMb{}{2}, \;
         \qMb{}{3}, \;
        -\qMb{}{4}, \;
        -\qMb{}{5}, \;
         \qMb{}{6}, \;
        -\qMb{}{7}, \;
         \qMb{}{8}  ]
    \end{equation}
Hermitian conjugation:
    \begin{equation}
        \qMb{q}{}^{\dag} =
       [ \qMb{}{1}^{T}, \;
        -\qMb{}{2}^{T}, \;
        -\qMb{}{3}^{T}, \;
        -\qMb{}{4}^{T}, \;
        -\qMb{}{5}^{T}, \;
         \qMb{}{6}^{T}, \;
         \qMb{}{7}^{T}, \;
         \qMb{}{8}^{T}  ].
    \end{equation}

The aforementioned simple operations, such as scaling, addition, and trace evaluation, are performed directly in biquaternion algebra and executed efficiently component-wise on the CPU using OpenMP parallelization~\cite{openmp}. The most computationally intensive operations, matrix multiplication and diagonalization, both of which exhibit formal cubic scaling with the matrix dimension, were selected as the principal targets for offloading to highly optimized external libraries, as they dominate the overall runtime of the library. On CPUs, \textsc{HMATLIB} utilizes Intel oneAPI MKL~\cite{intel_oneapi}, OpenBLAS bundled with the NVIDIA HPC SDK (hereafter NVHPC)~\cite{nvidia_hpcsdk} or NVPL (NVIDIA Performance Libraries)~\cite{nvidia_nvpl} designed by NVIDIA specifically for Arm $64$-bit architectures. In addition to these CPU backends, \textsc{HMATLIB} supports GPU acceleration through the cuBLAS and NVLAmath libraries provided by NVHPC.

For diagonalization, the backend LAPACK routines do not provide a dedicated optimized eigensolver for (bi)quaternion Hermitian matrices. Therefore, such matrices are first mapped onto their equivalent complex Hermitian representation using Eqs.~\eqref{eq:mapping-formulas} and subsequently diagonalized using \texttt{zheevd} for complex Hermitian matrices or, in scalar-relativistic calculations, \texttt{dsyevd} for real symmetric matrices. On CPUs, these routines are accessed directly through the standard LAPACK interface, whereas on GPUs, the same interface is accessed through the NVLAmath wrapper, which redirects the corresponding calls to the underlying C-based cuSOLVER routines~\cite{nvidia_fortran_cuda_2026}. In cases where the complex matrix possesses a specific block structure—either block-diagonal or block-antidiagonal—a structure-aware diagonalization based on a Hadamard-type unitary transformation is performed. Details of the procedure are described in Sec.~\ref{sec:implementation/diagonalization-hadamard}.

In contrast to diagonalization, \textsc{HMATLIB} performs matrix--matrix multiplication directly in the biquaternion algebra using the external \texttt{dgemm} routine on CPUs or \texttt{cublasDgemm} on GPUs. For the multiplication of two matrix-valued biquaternions with all components active (nonzero), a straightforward component-wise algorithm is substantially less efficient than the corresponding isomorphic complex matrix--matrix multiplication. To overcome this limitation, we developed and implemented the so-called \emph{bilinear biquaternion multiplication}, which extends the original approach for real quaternions~\cite{quatpdt} to the algebra of biquaternions and substantially reduces the computational cost of this operation. Details of the procedure are described in the following section.

\subsection{Bilinear Biquaternion Multiplication}
\label{sec:implementation/multiplication}
\noindent
The multiplication of two matrix-valued biquaternions, each having all eight components active ($[1,1,1,1,1,1,1,1]$), would require $64$ real matrix--matrix multiplications using a naive component-wise algorithm. As shown in the Results section, the computational time is considerably longer than that required for the corresponding isomorphic complex matrix--matrix multiplication. However, a significant acceleration of biquaternion multiplication can be achieved using our developed bilinear multiplication approach, which extends the method proposed by Howell and Lafon for real quaternions~\cite{quatpdt} and combines it with Gauss's three-multiplication algorithm for complex multiplication~\cite{knuth1997art}, reducing the total number of real matrix--matrix multiplications from $64$ to $24$.

The central idea of the bilinear biquaternion multiplication approach is the observation that the components $\mat{Q}_{1-4}\in\bbC^{m\times p}$ of the product
\begin{equation}
  \qQb{q}{}
  = \qMb{q}{} \qNb{q}{}
  = \mat{Q}_1\qI + \mat{Q}_2\qi + \mat{Q}_3\qj + \mat{Q}_4\qk
  \in\bbH_{\bbC}^{m \times p}
\end{equation}
 of two matrix-valued biquaternions
\begin{equation}
\begin{aligned}
\qMb{q}{} &= \mat{M}_1\qI + \mat{M}_2\qi + \mat{M}_3\qj + \mat{M}_4\qk \in\bbH_{\bbC}^{m \times n},
\\
\qNb{q}{} &= \mat{N}_1\qI\; + \mat{N}_2\qi + \mat{N}_3\qj\; + \mat{N}_4\qk\;
\in\bbH_{\bbC}^{n \times p},
\end{aligned}
\end{equation}
can be obtained as
\begin{equation}
\begin{aligned}
    \mat{Q}_1 &= ~~2\mat{P}_1 - \frac{1}{4}\left(\mat{P}_5+\mat{P}_6+\mat{P}_7+\mat{P}_8\right),
    \\
    \mat{Q}_2 &= -2\mat{P}_2 + \frac{1}{4}\left(\mat{P}_5+\mat{P}_6-\mat{P}_7-\mat{P}_8\right),
    \\
    \mat{Q}_3 &= -2\mat{P}_3 + \frac{1}{4}\left(\mat{P}_5-\mat{P}_6+\mat{P}_7-\mat{P}_8\right),
    \\
    \mat{Q}_4 &= -2\mat{P}_4 + \frac{1}{4}\left(\mat{P}_5-\mat{P}_6-\mat{P}_7+\mat{P}_8\right).
\end{aligned}
\end{equation}
Here, $\mat{P}_{1-8}\in\bbC^{m\times p}$ are intermediate matrices defined by the following eight complex matrix--matrix multiplications:
\begin{equation}
\begin{aligned}
    \mat{P}_1 &= \mat{M}_1\mat{N}_1,\;
    \mat{P}_2  = \mat{M}_4\mat{N}_3,\;
    \mat{P}_3  = \mat{M}_2\mat{N}_4,\;
    \mat{P}_4  = \mat{M}_3\mat{N}_2,
    \\
    \mat{P}_5 &= \left(\mat{M}_1+\mat{M}_2+\mat{M}_3+\mat{M}_4\right) \left(\mat{N}_1+\mat{N}_2+\mat{N}_3+\mat{N}_4\right),
    \\
    \mat{P}_6 &= \left(\mat{M}_1+\mat{M}_2-\mat{M}_3-\mat{M}_4\right) \left(\mat{N}_1+\mat{N}_2-\mat{N}_3-\mat{N}_4\right),
    \\
    \mat{P}_7 &= \left(\mat{M}_1-\mat{M}_2+\mat{M}_3-\mat{M}_4\right) \left(\mat{N}_1-\mat{N}_2+\mat{N}_3-\mat{N}_4\right),
    \\
    \mat{P}_8 &= \left(\mat{M}_1-\mat{M}_2-\mat{M}_3+\mat{M}_4\right) \left(\mat{N}_1-\mat{N}_2-\mat{N}_3+\mat{N}_4\right).
    \label{eq:bilinearH}
\end{aligned}
\end{equation}
Using conventional complex matrix multiplication, this approach requires $8\times4=32$ real matrix--matrix multiplications. This number can be further reduced to $24$ by evaluating each intermediate matrix in Eq.~\eqref{eq:bilinearH} using Gauss's complex multiplication scheme~\cite{knuth1997art}:
\begin{equation}
    \mat{P} = (\mat{T}_1 - \mat{T}_2) + i(\mat{T}_3 - \mat{T}_1 - \mat{T}_2). \label{cmplx-3-real-mult}
\end{equation}
where
\begin{equation}
\begin{aligned}
        \mat{T}_1 &= \mat{M}_{\text{R}}  \mat{N}_{\text{R}}, \quad
        \mat{T}_2 = \mat{M}_{\text{I}}  \mat{N}_{\text{I}}, \\
        \mat{T}_3 &= (\mat{M}_{\text{R}} + \mat{M}_{\text{I}})  (\mat{N}_{\text{R}} + \mat{N}_{\text{I}}).
    \end{aligned}
\end{equation}
In this form, each intermediate matrix requires only three real matrix--matrix multiplications, reducing the total number required by the bilinear biquaternion formulation to $8\times3=24$. It has been shown that no further reduction in the number of real multiplications is possible~\cite{winograd1971number}. For purely real ($[1,1,1,1,0,0,0,0]$) or purely imaginary ($[0,0,0,0,1,1,1,1]$) quaternion matrices, the bilinear formulation requires only $8$ real matrix--matrix multiplications, compared with $16$ for the naive component-wise algorithm. If some biquaternion components are absent, the corresponding terms can be omitted, further reducing the number of multiplications.

\subsection{Hadamard Transformation for Diagonalization}
\label{sec:implementation/diagonalization-hadamard}
\noindent
As discussed in Sec.~\ref{sec:implementation}, a Hermitian matrix-valued biquaternion is converted to its isomorphic complex representation and subsequently diagonalized. However, when some biquaternion components are absent, the complex matrix may acquire a block-diagonal or block-antidiagonal structure, allowing the eigenvalue problem to be reduced by exploiting this structure. This approach is both more memory efficient and significantly faster than diagonalizing the full complex matrix.

If the complex representation has a block-diagonal form
\begin{equation}
\mat{M} =
\begin{bmatrix}
    \mat{A} & \mat{0} \\
    \mat{0} & \mat{B}
\end{bmatrix}
\in \mathbb{C}^{2n \times 2n},
\end{equation}
the problem trivially reduces to at most two independent diagonalizations of Hermitian $n\times n$ matrices, with a further reduction possible if the two blocks are related by symmetry, as occurs in certain special cases. Conversely, when the complex representation yields a purely block anti-diagonal matrix
\begin{equation}
\mat{M} =
\begin{bmatrix}
    \mat{0} & \mat{A} \\
    \mat{B} & \mat{0}
\end{bmatrix}
\in \mathbb{C}^{2n \times 2n},
\label{eq:block-antidiagonal}
\end{equation}
a \emph{Hadamard-type unitary transformation} can be applied. The transformation yields a block-diagonal matrix whose upper and lower blocks differ only by a sign.

To demonstrate the approach, let us first consider a complex block anti-diagonal matrix of the form given in Eq.~\eqref{eq:block-antidiagonal}, with $n \times n$ subblocks $\mat{A}$ and $\mat{B}$ composed exclusively of pure $\qj$ ($\mat{F}_{\text{R}}+i\mat{H}_{\text{I}}$) or $\qk$ ($\mat{F}_{\text{I}}-i\mat{H}_{\text{R}}$) components in the sense of Eq.~\eqref{main-biquat-eqn}. The transformation of $\mat{M}$
\begin{equation}
    \mat{M}' = \mat{U}_{\alpha}^{\dag} \mat{M} \mat{U}_{\alpha}
\end{equation}
requires a simple Hadamard-type unitary matrix
\begin{equation}
\mat{U}_{\alpha} = \frac{1}{\sqrt{2}}
    \begin{bmatrix}
        \mat{I} & \mat{I} \\
        \alpha \mat{I} & -\alpha \mat{I}
    \end{bmatrix}
    \in \mathbb{C}^{2n \times 2n}
\end{equation}
yielding:
\begin{equation}
   \mat{M}' =
    \frac{1}{2}
    \begin{bmatrix}
        \alpha\mat{A}+\alpha^{*}\mat{B}
        &
        -\alpha\mat{A}+\alpha^{*}\mat{B}
        \\[4pt]
        \alpha\mat{A}-\alpha^{*}\mat{B}
        &
        -\alpha\mat{A}-\alpha^{*}\mat{B}
    \end{bmatrix}.
    \label{hadamard_general_transform}
\end{equation}
Here, $\alpha$ is a scalar satisfying $\alpha\alpha^{*} = |\alpha|^{2}=1$, as required by the unitarity of $\mat{U}_{\alpha}$. For $\mat{M}'$ to be block diagonal, it imposes $\alpha^{*}\mat{B} = \alpha\mat{A}$, which implies
\begin{equation}
    \mat{M}' =
    \begin{bmatrix}
        \alpha\mat{A} & \mat{0} \\
        \mat{0} & -\alpha\mat{A}
    \end{bmatrix}.
    \label{reduced-hadamard-pure}
\end{equation}
In the presence of $\qj$ component only, the conditions $\alpha^{*}\mat{B} = \alpha\mat{A}$ and $|\alpha|^{2}=1$ are satisfied for $\alpha=\pm i$. Hence,
\begin{equation}
    \mat{M}' =
    \begin{bmatrix}
        \pm i(\mat{F}_{\text{R}}+i\mat{H}_{\text{I}}) & \mat{0} \\
        \mat{0} & \mp i(\mat{F}_{\text{R}}+i\mat{H}_{\text{I}})
    \end{bmatrix}.
    \label{reduced-hadamard-pure}
\end{equation}
Similarly, in the presence of $\qk$ component only, the conditions $\alpha^{*}\mat{B} = \alpha\mat{A}$ and $|\alpha|^{2}=1$ are satisfied for $\alpha=\pm 1$. Hence,
\begin{equation}
    \mat{M}' =
    \begin{bmatrix}
        \pm (\mat{H}_{\text{R}}+i\mat{F}_{\text{I}}) & \mat{0} \\
        \mat{0} & \mp (\mat{H}_{\text{R}}+i\mat{F}_{\text{I}})
    \end{bmatrix}.
    \label{reduced-hadamard-pure}
\end{equation}

In the more general case where the complex block anti-diagonal matrix $\mat{M}$ contains contributions from both the $\qj$ and $\qk$ components, it is straightforward to show that the Hadamard-type unitary matrix can be written as
\begin{equation}
 \mat{U}_{\text{G}} = \frac{1}{\sqrt{2}}
 \begin{bmatrix}
    \mat{I} & \mat{I} \\
    \mat{G} & -\mat{G}
  \end{bmatrix}
  \in \mathbb{C}^{2n \times 2n}.
\end{equation}
For a matrix $\mat{A}$ with the singular value decomposition $\mat{A}=\mat{U}\mat{\Sigma}\mat{V}^{\dagger}$, the complex matrix $\mat{G}$, satisfying $\mat{G}^{\dagger}\mat{G}=\mat{I}$ and $\mat{A}\mat{G}=(\mat{A}\mat{G})^{\dagger}$, can be constructed from the left and right singular-vector matrices as $\mat{G}=\mat{V}\mat{U}^{\dagger}$.

\section{Results}
\label{sec:results}

\begin{table*}
\caption{\label{tab:specs}
Hardware and software environment of the tested GH200, H200 and A100 systems.
}
\begin{ruledtabular}
\begin{tabular}{llccc}
\textbf{Hardware} & \textbf{Details} & \textbf{GH200} & \textbf{H200} & \textbf{A100} \\
\hline
CPU Architecture & & ARM aarch64 & x86\_64 & x86\_64 \\
CPU Model & & NVIDIA Grace CPU & AMD EPYC 9654 & AMD EPYC 9654 \\
GPU Model & & NVIDIA GH200 & NVIDIA H200 NVL & NVIDIA A100 PCIe \\
CPU Cores & & 288 (72 cores/socket) & 96 (96 cores/socket) & 96 (96 cores/socket) \\
GPU Multiprocessors (SMs)
&
& $132$
& $132$
& $108$ \\
GPU FP64 Cores per SM
&
& $64$
& $64$
& $32$ \\
CPU Memory (in MiB) &  & 475\,136 & 386\,411 & 256\,888 \\
GPU Memory (in MiB) &  & 97\,871 & 143\,771 & 81\,920 \\
NVIDIA Driver & & 580.159.04 & 595.58.03 & 590.41.01 \\[0.1cm]

\textbf{Software} \\
\hline

Intel oneAPI\footnote{Intel oneAPI Toolkit~\cite{intel_oneapi}.}
& Version
& ---
& 2025.3.2\footnote{CPU compilation/linking: ifx -O3 -xHost -qopenmp -lmkl\_intel\_lp64 -lmkl\_intel\_thread -lmkl\_core -lpthread -lm -ldl.}
& 2022.2.1\footnote{CPU compilation/linking: ifx -O3 -xHost -qopenmp -lmkl\_intel\_lp64 -lmkl\_intel\_thread -lmkl\_core -lpthread -lm -ldl.}
\\

NVIDIA HPC SDK\footnote{NVHPC~\cite{nvidia_hpcsdk}.}
& Version (CUDA)

& 26.5 (13.2.0)\footnote{CPU compilation/linking: nvfortran -O3 -mp -tp=host
-lnvpl\_lapack\_lp64\_gomp -lnvpl\_blas\_lp64\_gomp.
\\\noindent
CPU/GPU compilation/linking: nvfortran -O3 -mp -tp=host -gpu=cc90,nvlamath -cuda -cudalib=cublas,nvlamath.
\\\noindent
Independently installed NVIDIA NVPL v26.5 was used rather than the NVPL library bundled with NVHPC; see Appendix~\ref{app-B:biquaternion-diagonalization} for details.}

& 26.3 (13.1.0)\footnote{CPU compilation/linking: nvfortran -O3 -mp -tp=host
-lblas\_lp64 -llapack\_lp64 -lm -ldl
\\\noindent
CPU/GPU compilation/linking: nvfortran -O3 -mp -tp=host -gpu=cc90,nvlamath -cuda -cudalib=cublas,nvlamath.}

& 25.3 (12.8.0)\footnote{CPU compilation/linking: nvfortran -O3 -mp -tp=host
-lblas\_lp64 -llapack\_lp64 -lm -ldl
\\\noindent
CPU/GPU compilation/linking: nvfortran -O3 -mp -tp=host -gpu=cc80,nvlamath -cuda -cudalib=cublas,nvlamath.}
\\

\end{tabular}
\end{ruledtabular}
\end{table*}

\noindent
Having established the theoretical background and implementation details of the biquaternion matrix library \textsc{HMATLIB}, we now assess its computational performance on modern CPU and GPU architectures. The assessment was carried out on three different systems, each equipped with a single NVIDIA A100, H200, or GH200 GPU. The different hardware and software configurations, summarized in Table~\ref{tab:specs}, allow us to determine whether the general performance trends are preserved across the systems. This section focuses on results obtained on the H200 system, while the corresponding results for the GH200 and A100 systems are provided in the Appendix.

The computational performance is assessed using the most computationally intensive operations, namely matrix multiplication and diagonalization, which constitute the principal computational bottlenecks in \textsc{HMATLIB}. The input matrices (either general or Hermitian) were generated in host (CPU) memory using a seeded pseudo-random number generator to ensure reproducibility of the input data. Subsequent calculations utilized all available CPU cores through the OpenMP parallel programming interface or, for GPU-accelerated calculations, all GPU cores through NVIDIA's CUDA parallel-computing platform.
The wall-clock time was measured using the \texttt{omp\_get\_wtime()} function. For GPU calculations, the reported timings include the transfer of input matrices from host to device memory, the GPU computation, and the transfer of the resulting matrices back to host memory, with \texttt{cudaDeviceSynchronize()} used to ensure completion of all GPU operations before stopping the timer~\cite{cudafortran}.

\subsection{Multiplication}
\label{sec:results/multiplication}

\begin{figure*}[!t]
\centering
\begin{minipage}{0.48\textwidth}
  \centering
  \fbox{%
  \begin{minipage}{0.97\linewidth}
    \centering
    \textbf{[1,1,1,1,1,1,1,1]}\\[0.5em]

    \includegraphics[width=\linewidth]{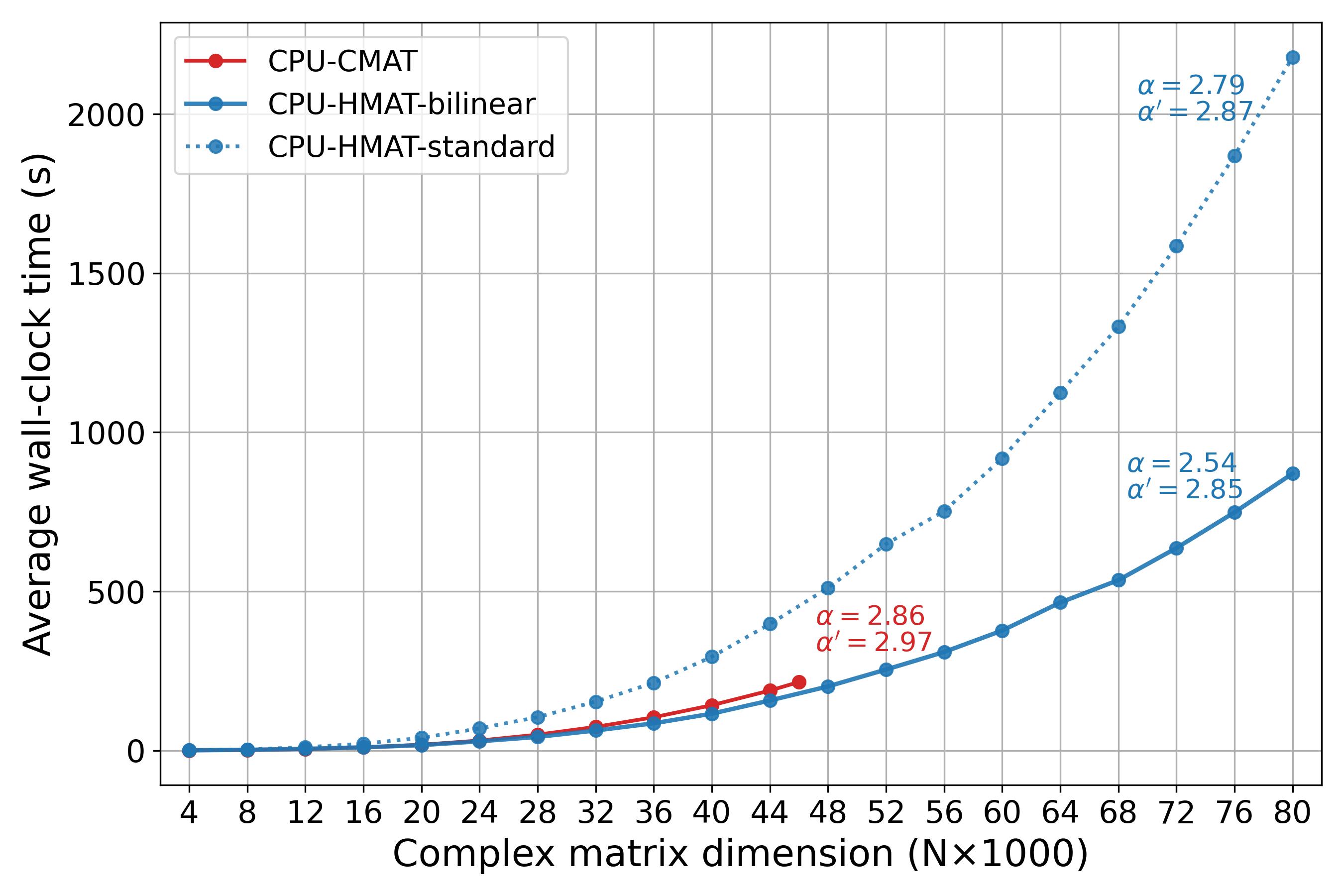}\\
    (a) Pure CPU

    \vspace{0.5em}

    \includegraphics[width=\linewidth]{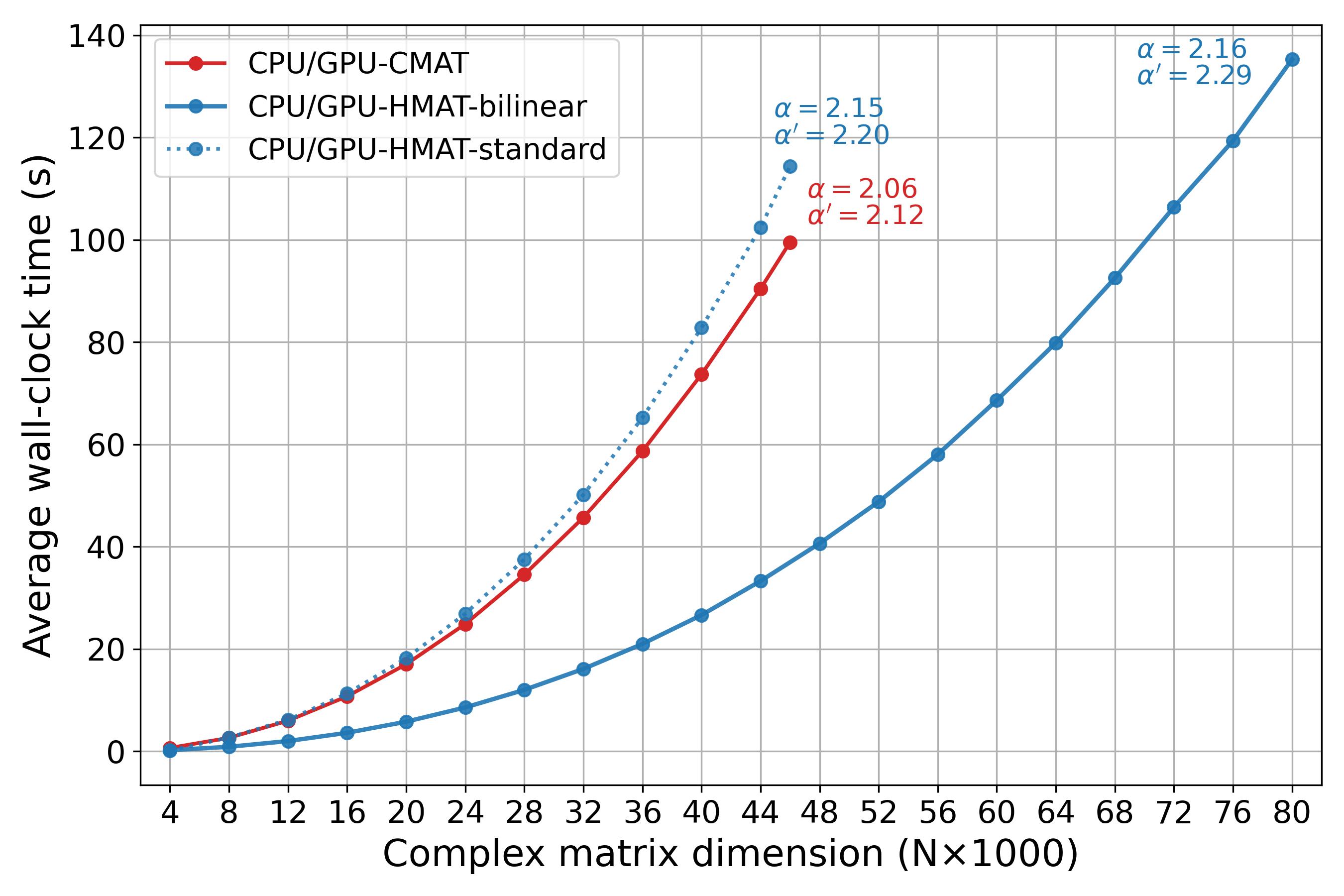}\\
    (b) Hybrid CPU/GPU
  \end{minipage}
  }
\end{minipage}
\hfill
\begin{minipage}{0.48\textwidth}
  \centering
  \fbox{%
  \begin{minipage}{0.97\linewidth}
    \centering
    \textbf{[1,1,1,1,0,0,0,0]}\\[0.5em]

    \includegraphics[width=\linewidth]{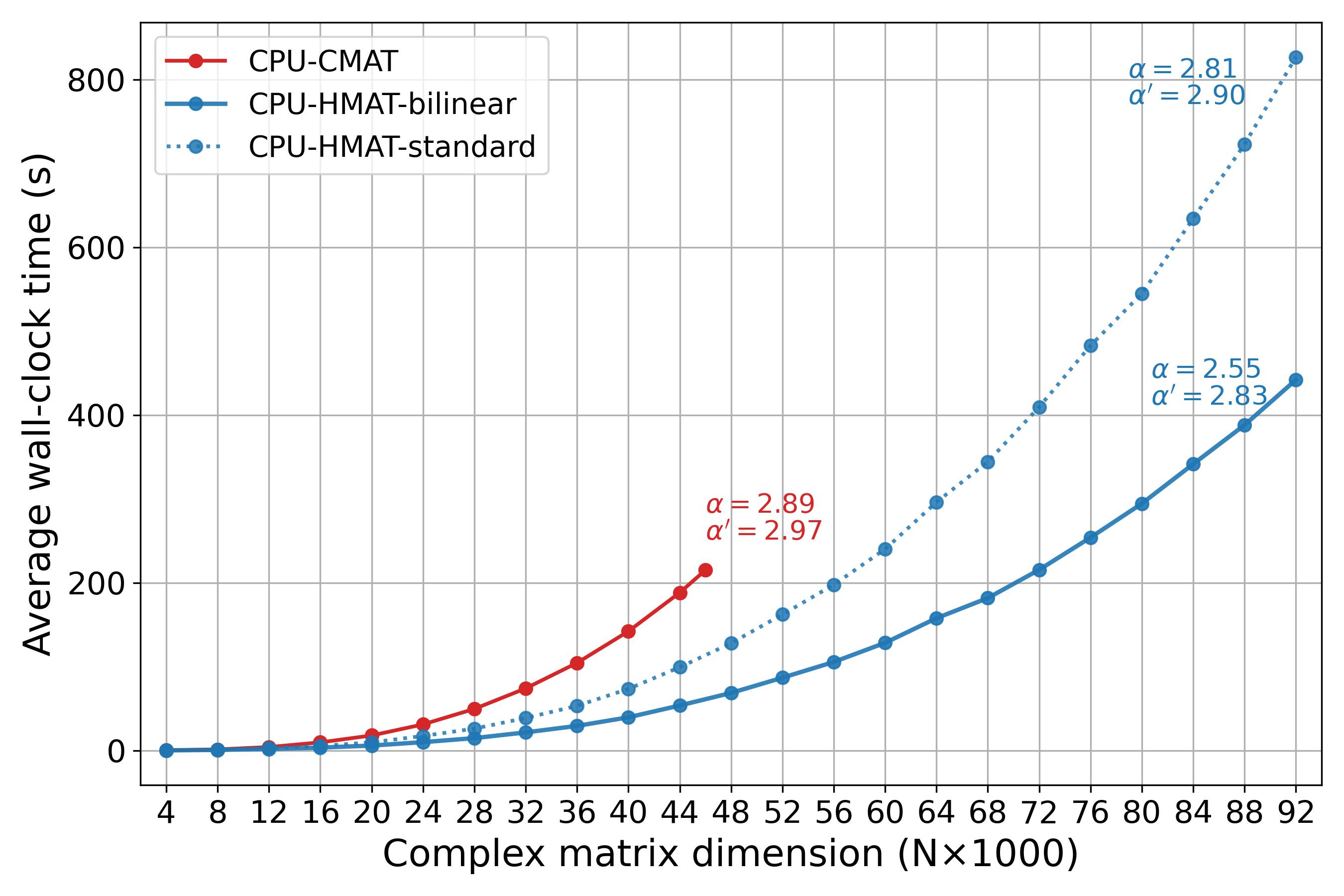}\\
    (c) Pure CPU

    \vspace{0.5em}

    \includegraphics[width=\linewidth]{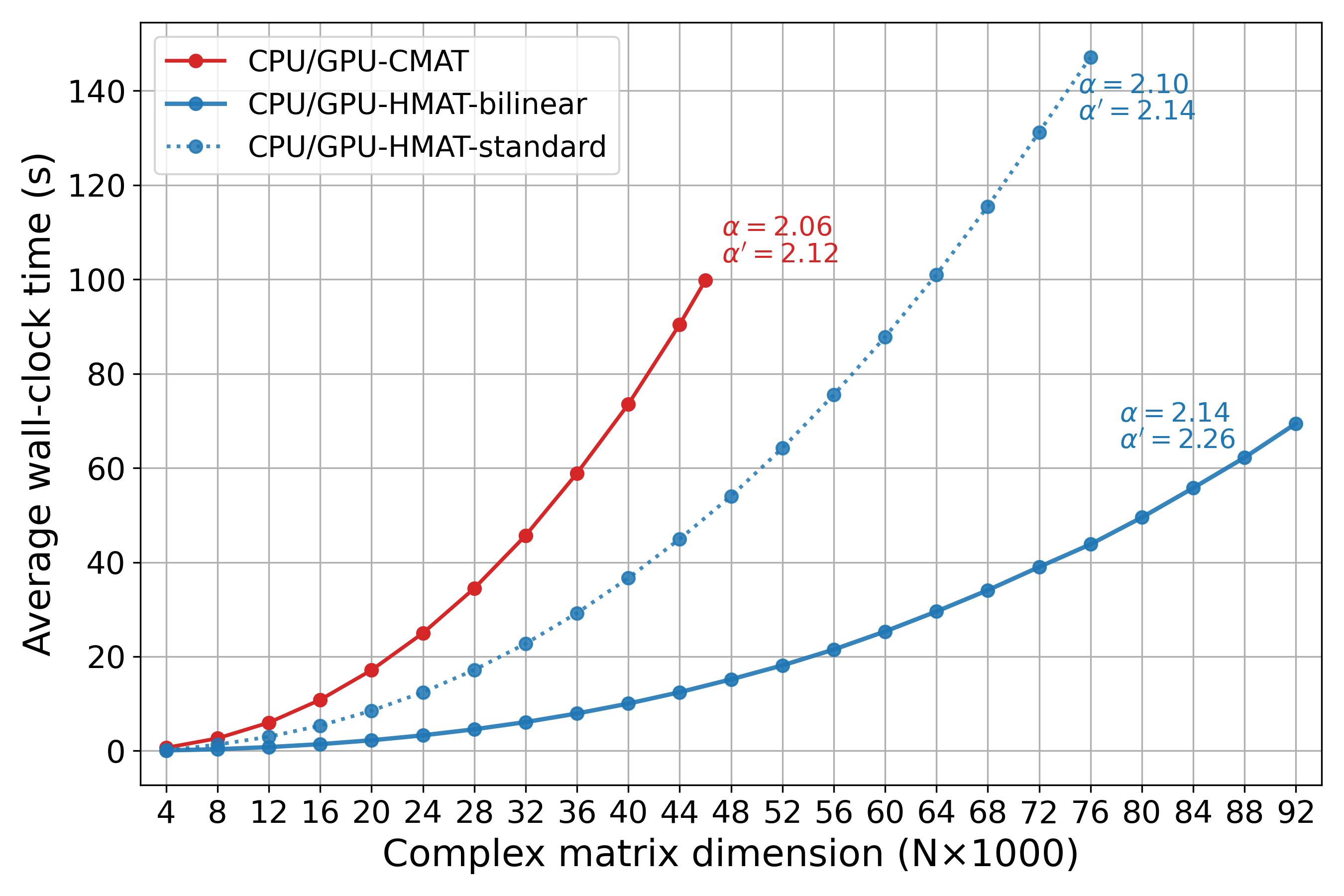}\\
    (d) Hybrid CPU/GPU
  \end{minipage}
  }
\end{minipage}
\caption{Average wall-clock time for complex (\textsc{CMAT}), standard biquaternion (\textsc{HMAT}-standard), and bilinear biquaternion (\textsc{HMAT}-bilinear) matrix--matrix multiplication as a function of the complex matrix dimension $N$. The left and right panels show the fully populated biquaternion ($[1,1,1,1,1,1,1,1]$) and real-quaternion ($[1,1,1,1,0,0,0,0]$) cases, respectively, with panels (a) and (c) corresponding to CPU-only calculations and panels (b) and (d) to hybrid CPU/GPU calculations. Wall-clock times were fitted to a power law, yielding the effective exponents $\alpha$ and $\alpha'$ over the complete range of $N$ and for $N\geq 20000$, respectively. All results were obtained on the H200 system using the \textsc{HMATLIB} library linked with the NVIDIA software libraries specified in Table~\ref{tab:specs}. The corresponding results for GH200 and A100 systems are given in Appendix~\ref{app-A:biquaternion-multiplication}.}
\label{h200-mult}
\end{figure*}

\begin{figure*}[!t]
    \centering
    \includegraphics[width=0.8\textwidth]{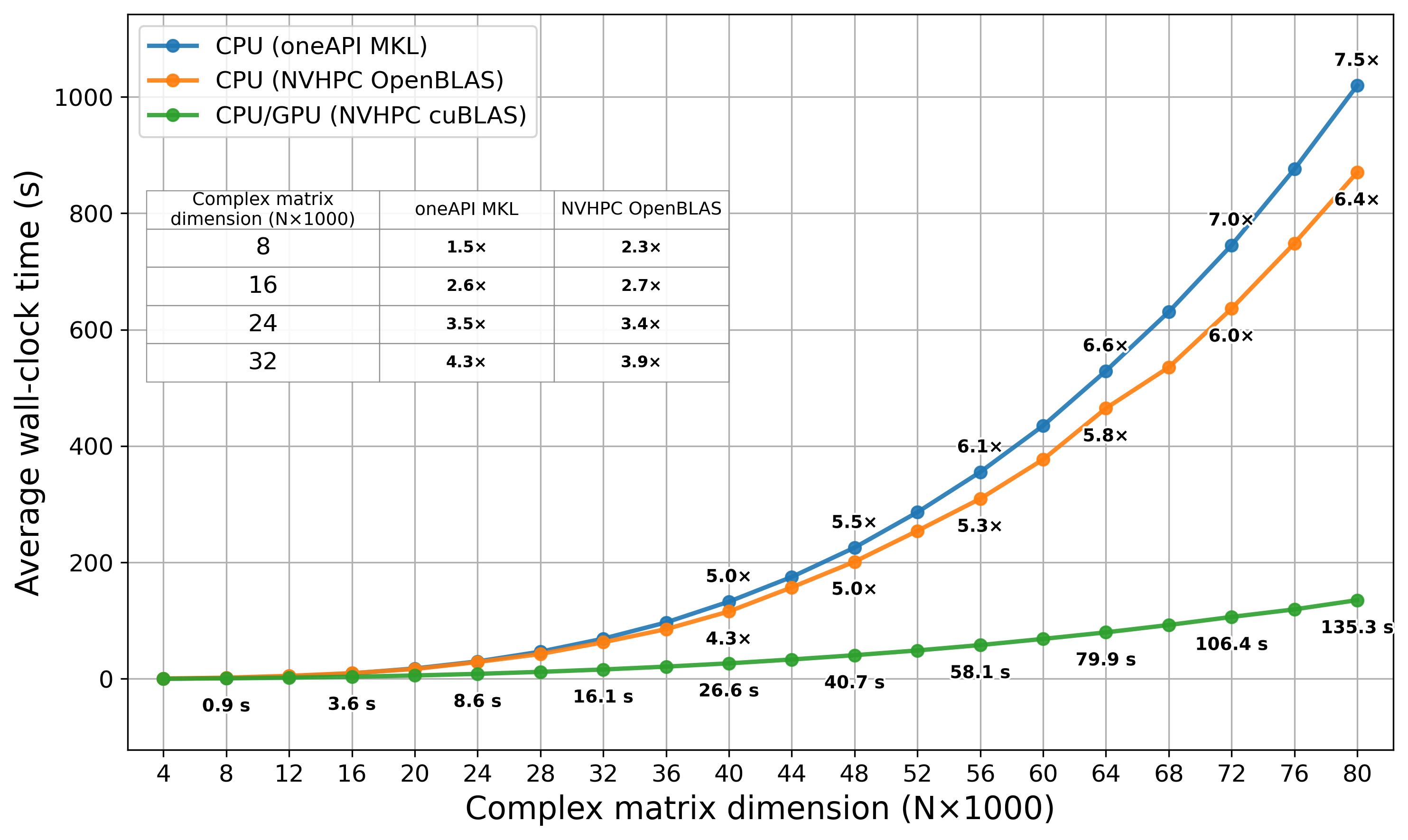}
    \caption{Average wall-clock time for bilinear biquaternion (\textsc{HMAT}-bilinear) matrix--matrix multiplication as a function of the complex matrix dimension $N$ for the fully populated biquaternion case ($[1,1,1,1,1,1,1,1]$). Pure CPU calculations using the oneAPI MKL and NVHPC OpenBLAS backends are compared with hybrid CPU/GPU calculations with the NVHPC cuBLAS backend. Values below the green curve report the absolute CPU/GPU execution times in seconds, while the adjacent labels to the CPU curves and the inset table report the ratio between CPU and CPU/GPU times. All results were obtained on the H200 system specified in Table~\ref{tab:specs}.}
    \label{h200-mult-cpu-gpu-compare-libs}
\end{figure*}

\noindent
To demonstrate the substantial performance gain achieved by the novel bilinear biquaternion matrix--matrix multiplication introduced in Section~\ref{sec:implementation/multiplication}, we compare the bilinear approach (\textsc{HMAT}-bilinear) with the standard component-wise biquaternion multiplication (\textsc{HMAT}-standard) and with the reference isomorphic complex matrix multiplication (\textsc{CMAT}) over a broad range of matrix sizes, from $4000\,\times\,4000$ to $92000\,\times\,92000$. The results are shown in Fig.~\ref{h200-mult} for the multiplication of two square matrix-valued biquaternions ($[1,1,1,1,1,1,1,1]$) and real quaternions ($[1,1,1,1,0,0,0,0]$). Each data point represents the wall-clock time averaged over all combinations of the transformations $T$, $*$, and $\dag$ applied independently to the two input matrices. CPU-only calculations are denoted by the prefix CPU, whereas GPU-accelerated calculations are denoted by the prefix CPU/GPU. The latter designation reflects the fact that the input matrices initially reside in host (CPU) memory and are transferred to device (GPU) memory for computation, with the resulting matrices transferred back to host memory upon completion.

In the case of full biquaternion ($[1,1,1,1,1,1,1,1]$) matrix--matrix multiplication [Figs.~\ref{h200-mult}(a) and (b)], \textsc{HMAT}-bilinear yields the shortest times in both CPU and hybrid CPU/GPU calculations. This performance gain can be attributed to the fact that the \textsc{HMAT}-bilinear approach requires only $24$ real matrix--matrix multiplications, compared with $64$ for the conventional component-wise approach (\textsc{HMAT}-standard), as discussed in Sec.~\ref{sec:implementation/multiplication}. This substantial reduction in the number of real matrix multiplications is sufficient for \textsc{HMAT}-bilinear to outperform even the isomorphic complex multiplication, \textsc{CMAT}. While the performance advantage of \textsc{HMAT}-bilinear is relatively modest for CPU calculations, it becomes considerably larger with hybrid CPU/GPU execution. For example, at a complex matrix dimension of $N=44000$, \textsc{HMAT}-bilinear requires approximately $157$ seconds on the CPU, corresponding to speedup factors of approximately $1.2$ and $2.5$ relative to \textsc{CMAT} and \textsc{HMAT}-standard, respectively [Fig.~\ref{h200-mult}(a)]. With hybrid CPU/GPU execution, the wall-clock time of \textsc{HMAT}-bilinear decreases to approximately $33$ seconds, while the corresponding speedup factors increase to $2.7$ and $3.1$, respectively [Fig.~\ref{h200-mult}(b)].

The computational advantage of biquaternion algebra over complex algebra becomes even more pronounced when fewer biquaternion components are present. For real-quaternion ($[1,1,1,1,0,0,0,0]$) matrix--matrix multiplication [Figs.~\ref{h200-mult}(c) and (d)], the computation time for \textsc{CMAT} remains the same as in the full-biquaternion $[1,1,1,1,1,1,1,1]$  case, whereas it decreases significantly for the biquaternion approaches. As a result, even \textsc{HMAT}-standard becomes faster than \textsc{CMAT}, while \textsc{HMAT}-bilinear outperforms \textsc{HMAT}-standard for both CPU and hybrid CPU/GPU execution, as \textsc{HMAT}-bilinear requires only 8 real matrix--matrix multiplications compared with 16 for \textsc{HMAT}-standard. For example, at the same complex matrix dimension of $N=44000$ discussed earlier, \textsc{HMAT}-bilinear requires approximately $54$ seconds on the CPU, corresponding to speedup factors of approximately $3.5$ and $1.9$ relative to \textsc{CMAT} and \textsc{HMAT}-standard, respectively [Fig.~\ref{h200-mult}(c)]. With hybrid CPU/GPU execution, the wall-clock time of \textsc{HMAT}-bilinear decreases to approximately $12$ seconds, while the corresponding speedup factors increase to $7.3$ and $3.6$, respectively [Fig.~\ref{h200-mult}(d)]. Overall, these results demonstrate a clear computational advantage of the \textsc{HMAT}-bilinear approach to matrix--matrix multiplication over the conventional complex-algebra formulation, with particularly large performance gains for real-quaternion matrices and hybrid CPU/GPU execution. The same trends are observed on the GH200 and A100 systems, with the corresponding results provided in Appendix~\ref{app-A:biquaternion-multiplication}.

The measured wall-clock times in Fig.~\ref{h200-mult} exhibit distinct scaling behavior for CPU-only and hybrid CPU/GPU calculations. For CPU-only calculations, power-law fits yield effective exponents $\alpha=2.54$--$2.89$ over the complete range of $N$ and $\alpha'=2.83$--$2.97$ for $N\geq20000$, indicating an approach toward the formal $O(N^3)$ scaling of dense matrix--matrix multiplication at larger matrix dimensions. In contrast, hybrid CPU/GPU calculations exhibit effectively subcubic scaling over the investigated range, with substantially smaller exponents of $\alpha=2.06$--$2.16$ and $\alpha'=2.12$--$2.29$, reflecting increasing GPU computational efficiency with $N$. In both cases, $\alpha'>\alpha$, indicating a gradual approach toward the asymptotic regime as $N$ increases. Finally, the similar exponents for the fully populated biquaternion and real-quaternion cases indicate that reducing the number of component multiplications primarily affects the computational prefactor rather than the scaling behavior.

Beyond the reduction in wall-clock time, the biquaternion formulation offers an additional computational advantage by allowing larger matrix dimensions to be treated. In our \textsc{CMAT} implementation, the elements of a complex matrix are stored in a one-dimensional array and indexed using $32$-bit integer. Consequently, the number of matrix elements must not exceed $2^{31}-1$, corresponding to a maximum square-matrix dimension $N$ of approximately $N\simeq 46000$. In \textsc{HMAT}, the same indexing constraint applies to the real component matrices of dimension $N/2$, effectively \emph{doubling} the maximum matrix dimension to approximately $N=92000$ [Figs.~\ref{h200-mult}(c) and (d)]. For the full-biquaternion case [Figs.~\ref{h200-mult}(a) and (b)], the largest matrix dimension attained with \textsc{HMAT}-bilinear was $N=80000$. This limit, however, is imposed by the available CPU memory on the H200 system rather than by $32$-bit indexing and therefore does not represent an intrinsic limitation of the biquaternion formulation.

For hybrid CPU/GPU calculations, the maximum accessible matrix dimension is further affected by the different device-memory management strategies employed by the two \textsc{HMAT} multiplication schemes. In \textsc{HMAT}-standard, all active input and product components are allocated simultaneously in device memory, thereby reducing repeated host--device transfers at the expense of a higher peak device-memory requirement. In contrast, \textsc{HMAT}-bilinear allows the operands to be transferred to and processed on the device one at a time, without the need to keep all components simultaneously in device memory. This substantially reduces the peak device-memory requirement, making the maximum accessible matrix dimension primarily limited by the available host memory. This difference in device-memory management explains why, for hybrid CPU/GPU calculations, the maximum matrix dimension attainable with \textsc{HMAT}-standard is consistently smaller than that attainable with \textsc{HMAT}-bilinear.

Finally, Fig.~\ref{h200-mult-cpu-gpu-compare-libs} summarizes the performance gain obtained by moving from pure CPU to hybrid CPU/GPU execution for \textsc{HMAT}-bilinear matrix--matrix multiplication. Two CPU backends, oneAPI MKL and NVHPC OpenBLAS, were considered and compared with CPU/GPU with the cuBLAS backend. The two CPU setups exhibit comparable performance at intermediate matrix dimensions, whereas NVHPC OpenBLAS becomes faster than oneAPI MKL as $N$ increases. The advantage of GPU acceleration also grows with matrix dimension. At $N=8000$, CPU/GPU provides moderate speedups of $1.5$ and $2.3$ relative to oneAPI MKL and NVHPC OpenBLAS, respectively, increasing to $7.5$ and $6.4$ at $N=80000$, where the hybrid CPU/GPU wall-clock time is only $135.3$~s. These results demonstrate that the computational advantage of bilinear biquaternion matrix--matrix multiplication can be further amplified by GPU acceleration, with increasingly pronounced performance gains at larger matrix dimensions.

\subsection{Diagonalization}
\label{sec:results/diagonalization}

\begin{figure*}[!t]
    \centering
    \includegraphics[width=0.8\textwidth]{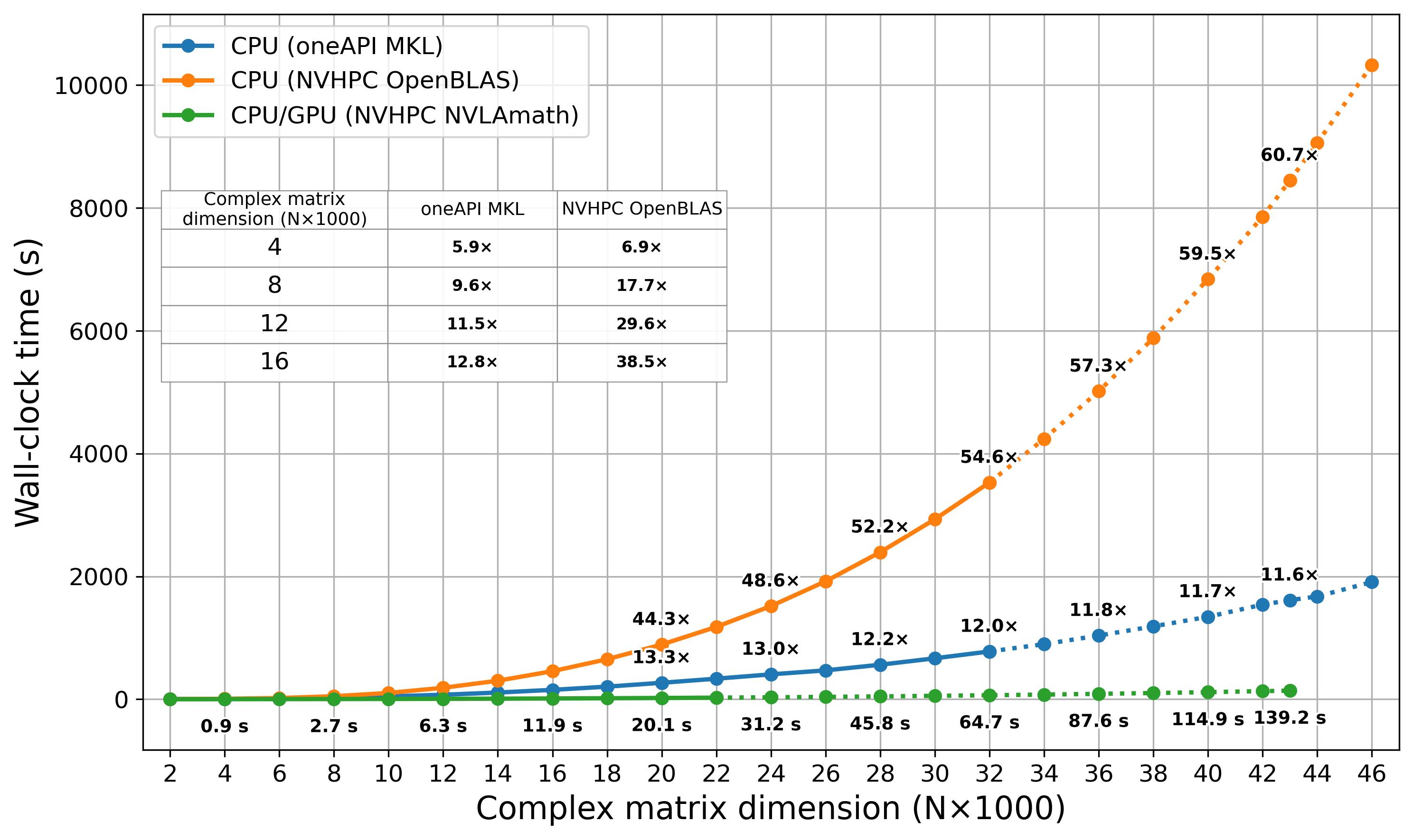}
    \caption{Wall-clock time for \textsc{HMATLIB} diagonalization as a function of the complex matrix dimension $N$. Pure CPU calculations using oneAPI MKL and NVHPC OpenBLAS backends are compared with hybrid CPU/GPU calculations using NVHPC NVLAmath backend. Values below the green curve report the absolute CPU/GPU execution times in seconds, while the adjacent labels to the CPU curves and the inset table report the ratio between CPU and CPU/GPU times.
    The solid curve segments indicate calculations performed with the $32$-bit integer interface while the dotted curve segments correspond to the $64$-bit interface. All calculations were performed on the H200 system using the software and hardware environments described in Table~\ref{tab:specs}. The corresponding results for GH200 and A100 systems are given in Appendix~\ref{app-B:biquaternion-diagonalization}.}
    \label{h200-diag-cpu-gpu-compare-libs}
\end{figure*}

\noindent
We next consider the second major computational bottleneck in \textsc{HMATLIB}, namely matrix diagonalization. For its assessment, we used fully populated square matrix-valued biquaternions ($[1,1,1,1,1,1,1,1]$) as inputs. In the current version of \textsc{HMATLIB}, these are mapped isomorphically to the corresponding complex Hermitian matrix representation, as discussed in Sec.~\ref{sec:implementation}. This complex representation enables the diagonalization to be offloaded to highly optimized external libraries. Preliminary benchmarks showed that the divide-and-conquer eigensolver \texttt{zheevd} consistently outperformed the alternative \texttt{zheev}, \texttt{zheevx}, and \texttt{zheevr} routines on the CPU, while the corresponding GPU eigensolvers exhibited comparable performance. Accordingly, \texttt{zheevd} was used as the backend routine for both CPU and CPU/GPU matrix diagonalizations in \textsc{HMATLIB}. Figure~\ref{h200-diag-cpu-gpu-compare-libs} compares their performance over a range of matrix dimensions using the respective numerical backends.

As expected, the hybrid CPU/GPU implementation backed by the NVLAmath library yields the shortest wall-clock times over the entire range of matrix dimensions considered. For instance, at $N=4000$, the hybrid implementation is approximately $6$ times faster than the CPU implementation using oneAPI MKL and $7$ times faster than that using NVHPC OpenBLAS. At $N=24000$, the corresponding speedups increase to $13$ and $49$ times, respectively. At $N=43000$, the maximum matrix dimension accommodated by the available device memory, the hybrid implementation is approximately $12$ times faster than oneAPI MKL and $61$ times faster than NVHPC OpenBLAS. Of the two CPU backends, oneAPI MKL yields considerably shorter wall-clock times than NVHPC OpenBLAS for matrix diagonalization. This behavior contrasts with the multiplication results, where NVHPC OpenBLAS becomes faster than oneAPI MKL at large matrix dimensions (see Fig.~\ref{h200-mult-cpu-gpu-compare-libs}).

Note that the CPU/GPU timings correspond to the complete algorithmic workflow rather than to the eigensolver alone. This includes device-memory allocation and deallocation, host--device data transfers, mapping of the matrix-valued biquaternion to its isomorphic complex Hermitian matrix representation, Hermiticity checks, workspace queries, and eigensolver execution. A breakdown of the total CPU/GPU wall-clock time is given in Table~\ref{tab:h200-diag-transfer}. The host--device data transfers account for approximately $46\%$ of the total time at $N=8000$, decreasing to $36\%$ at $N=24000$ and $26\%$ at $N=43000$. Correspondingly, the eigensolver accounts for an increasing fraction of the total wall-clock time as the matrix dimension increases. Despite the additional mapping and data-transfer overhead, the complete CPU/GPU execution remains substantially faster than pure CPU execution.

The use of \texttt{zheevd} introduces an additional limitation associated with its workspace requirements. For larger matrices, the required workspace length may exceed the range of a signed $32$-bit integer even when the matrix dimension itself remains within this range. Our CPU implementations therefore switch from the $32$-bit to the $64$-bit eigensolver interface at approximately $N=32760$, where the real workspace length approaches the $2^{31}-1$ limit. For the NVLAmath backend, the corresponding transition occurs earlier, at approximately $N=23170$, beyond which the complex workspace length exceeds the $32$-bit indexing limit. These transitions are indicated by the change from solid to dotted curve segments in Fig.~\ref{h200-diag-cpu-gpu-compare-libs}. The appropriate interface is selected automatically based on the matrix dimension, allowing the same \texttt{zheevd} algorithm to be used beyond the $32$-bit workspace limit.

\begin{table}
\caption{\label{tab:h200-diag-transfer}
Breakdown of the CPU/GPU wall-clock time into eigensolver computation and data transfers for \textsc{HMATLIB} diagonalization on the H200 system.}
\begin{ruledtabular}
\begin{tabular}{cccc}
& \multicolumn{3}{c}{Time (s)} \\
\cline{2-4}
\textit{N} &
Total &
Eigensolver &
Data Transfers \\
\hline
8000  & 2.733   & 1.077   & 1.248  \\
16000 & 11.867  & 6.263   & 4.948  \\
24000 & 31.237  & 19.015  & 11.128 \\
32000 & 64.678  & 43.288  & 19.651 \\
40000 & 114.937 & 81.708  & 30.654 \\
43000 & 139.173 & 100.884 & 35.426 \\
\end{tabular}
\end{ruledtabular}
\end{table}

\section{Conclusion}
\label{sec:conclusion}
\noindent

In this work, we have discussed biquaternion algebra as a unified mathematical framework for relativistic electronic-structure theory. Starting from the connection between time-reversal symmetry and quaternion algebra, we showed that the complementary time-reversal-antisymmetric structures extend this representation to complex quaternions, or biquaternions. We have explicitly exploited biquaternion algebra for fundamental objects in relativistic electronic-structure theory, including operators, kinetically and magnetically balanced basis functions, and the corresponding
expectation values. Since the algebras of real numbers, complex numbers, and real quaternions are naturally embedded as subalgebras of the biquaternion algebra, the resulting formulation provides a common algebraic framework for relativistic, scalar-relativistic, and nonrelativistic electronic-structure theory.

To realize this framework computationally, we have developed \textsc{HMATLIB}, a biquaternion matrix library implemented within the \textsc{ReSpect} DFT package for both pure CPU and hybrid CPU/GPU execution. Its core \textsc{HMAT} module represents a matrix-valued biquaternion through eight real component matrices and performs simple operations directly on these components using OpenMP parallelization, while the computationally demanding matrix multiplication and diagonalization are delegated to highly optimized numerical backends. A central development is the bilinear biquaternion matrix multiplication algorithm, which combines a bilinear quaternion product with Gauss's three-multiplication scheme for complex numbers and reduces the number of real matrix--matrix multiplications from $64$ to $24$ for fully populated biquaternions and from $16$ to $8$ for real quaternions. For diagonalization, matrix-valued biquaternions are mapped isomorphically to complex Hermitian matrices, allowing the use of highly optimized \texttt{zheevd} eigensolvers. The library supports several CPU backends as well as GPU acceleration through cuBLAS and NVLAmath, providing a common implementation for modern CPU and heterogeneous CPU/GPU architectures.

The numerical benchmarks on the H200 system, supported by corresponding results on the GH200 and A100 systems, demonstrate that these algebraic and algorithmic advantages translate directly into substantial performance gains. For fully populated biquaternion multiplication at $N=44000$, \textsc{HMAT}-bilinear is approximately $1.2$ times faster than \textsc{CMAT} on the CPU, with the speedup increasing to $2.7$ times for hybrid CPU/GPU execution. For real-quaternion matrices, the corresponding speedups increase to $3.5$ and $7.3$ times, respectively. Moreover, the component representation in \textsc{HMAT} effectively doubles the maximum matrix dimension imposed by $32$-bit indexing from approximately $N=46000$ to $N=92000$. Matrix diagonalization benefits even more strongly from GPU acceleration: at $N=43000$, the complete CPU/GPU workflow is approximately $12$ times faster than oneAPI MKL and $61$ times faster than NVHPC OpenBLAS, despite including the biquaternion-to-complex mapping and host--device data transfers. Overall, these results demonstrate that biquaternion algebra provides not only a natural and unified representation of relativistic electronic-structure quantities, but also a computationally advantageous foundation for large-scale calculations on modern CPU and GPU architectures.

\begin{acknowledgments}
\noindent
We thank Rasmus Vikhamar-Sandberg for drawing our attention to the bilinear approach. We acknowledge support from the Research Council of Norway through its Centres
of Excellence scheme (project No.~262695) and research grant No.~315822. M.R.
also acknowledges funding from the EU NextGenerationEU through the Recovery and
Resilience Plan for Slovakia under the project No.~09I05-03-V02-00034, as well
as from the Slovak Research and Development Agency (grants No.~APVV-25-0750 and
DS-FR-24-0045) and VEGA (grant No.~1/0670/24). T. S. acknowledges funding by the European Research Council (ERC) under the European Union’s Horizon 2020 research and innovation programme (Grant Agreement No. ID:101019907). Part of the research results
were obtained using the computational resources funded by the EU
NextGenerationEU through the Recovery and Resilience Plan for Slovakia under
the project No.~17I03-04-P02-00001, Development and construction of a
supercomputer for national supercomputing centre.
\end{acknowledgments}

\bibliography{references}

\appendix

\section{\textsc{HMATLIB} Multiplication}
\label{app-A:biquaternion-multiplication}

\begin{figure*}[!t]
\centering
\begin{minipage}{0.48\textwidth}
  \centering
  \fbox{%
  \begin{minipage}{0.97\linewidth}
    \centering
    \textbf{[1,1,1,1,1,1,1,1]}\\[0.5em]

    \includegraphics[width=\linewidth]{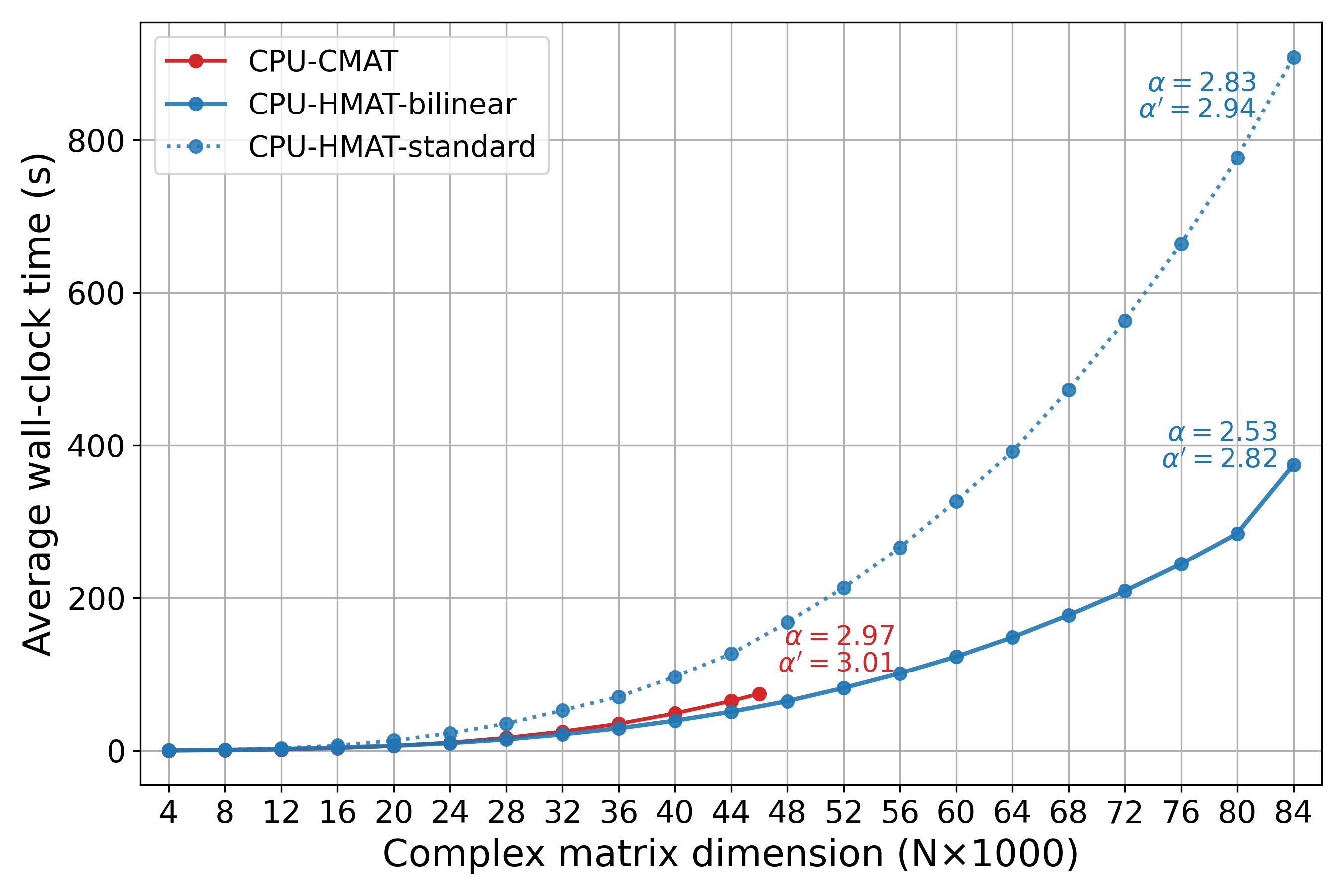}\\
    (a) Pure CPU

    \vspace{0.5em}

    \includegraphics[width=\linewidth]{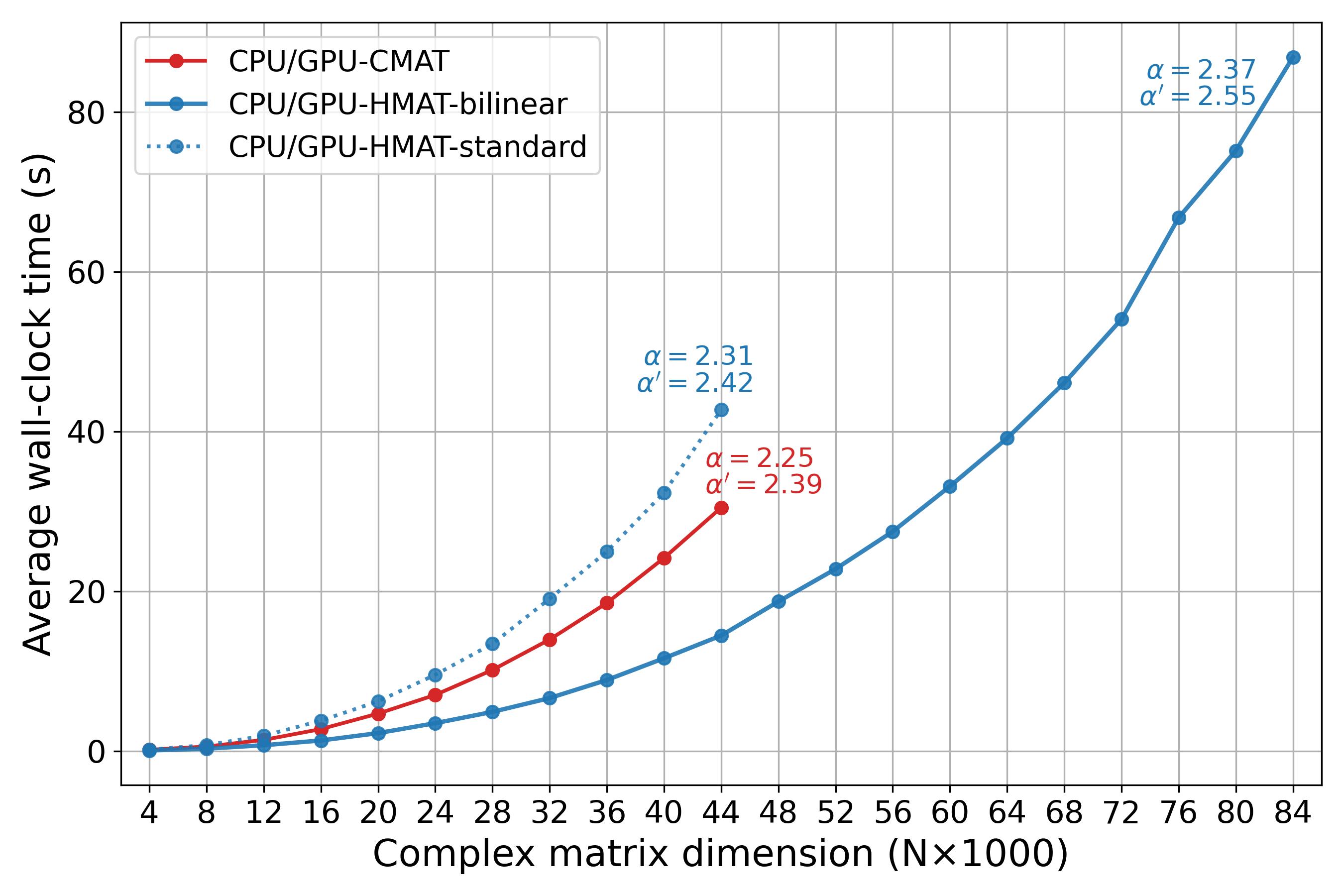}\\
    (b) Hybrid CPU/GPU
  \end{minipage}
  }
\end{minipage}
\hfill
\begin{minipage}{0.48\textwidth}
  \centering
  \fbox{%
  \begin{minipage}{0.97\linewidth}
    \centering
    \textbf{[1,1,1,1,0,0,0,0]}\\[0.5em]

    \includegraphics[width=\linewidth]{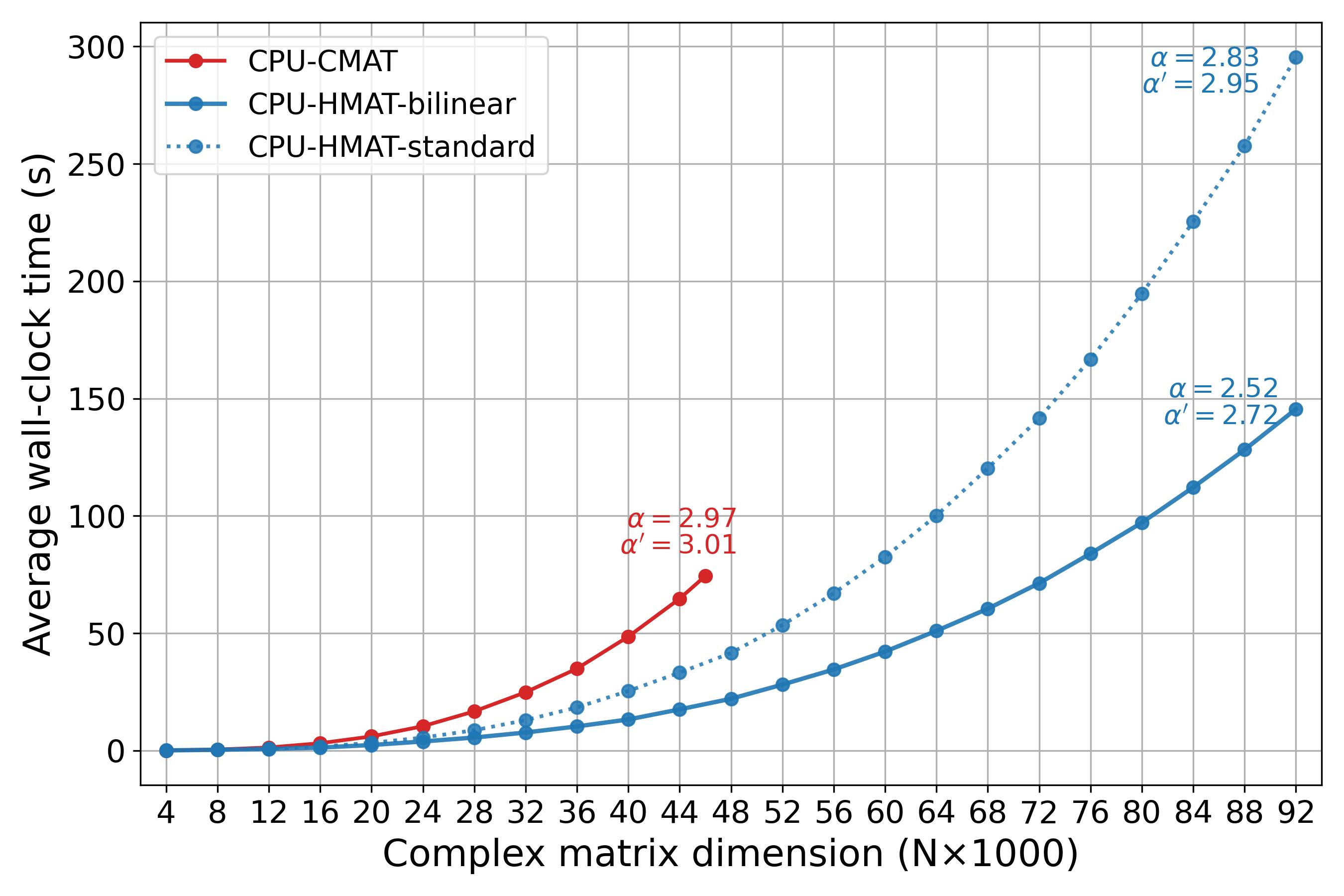}\\
    (c) Pure CPU

    \vspace{0.5em}

    \includegraphics[width=\linewidth]{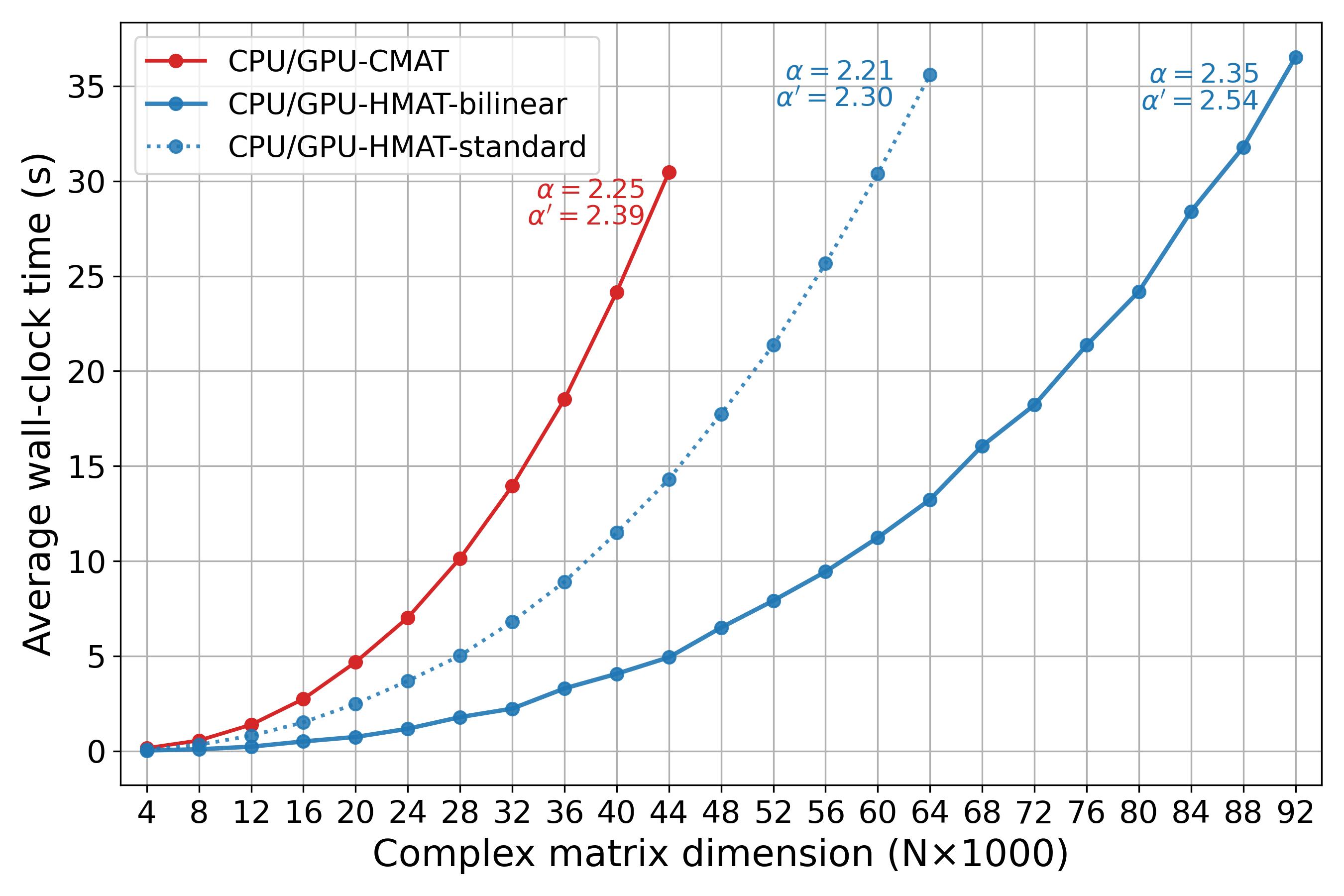}\\
    (d) Hybrid CPU/GPU
  \end{minipage}
  }
\end{minipage}
\caption{Average wall-clock time for complex (\textsc{CMAT}), standard biquaternion (\textsc{HMAT}-standard), and bilinear biquaternion (\textsc{HMAT}-bilinear) matrix--matrix multiplication as a function of the complex matrix dimension $N$. The left and right panels show the fully populated biquaternion ($[1,1,1,1,1,1,1,1]$) and real-quaternion ($[1,1,1,1,0,0,0,0]$) cases, respectively, with panels (a) and (c) corresponding to CPU-only calculations and panels (b) and (d) to hybrid CPU/GPU calculations. Wall-clock times were fitted to a power law, yielding the effective exponents $\alpha$ and $\alpha'$ over the complete range of $N$ and for $N\geq 20000$, respectively. All results were obtained on the \textbf{GH200 system} using the \textsc{HMATLIB} library linked with the NVIDIA software libraries specified in Table~\ref{tab:specs}.}
\label{gh200-mult-2}
\end{figure*}

\begin{figure*}[!t]
\centering
\begin{minipage}{0.48\textwidth}
  \centering
  \fbox{%
  \begin{minipage}{0.97\linewidth}
    \centering
    \textbf{[1,1,1,1,1,1,1,1]}\\[0.5em]
    \includegraphics[width=\linewidth]{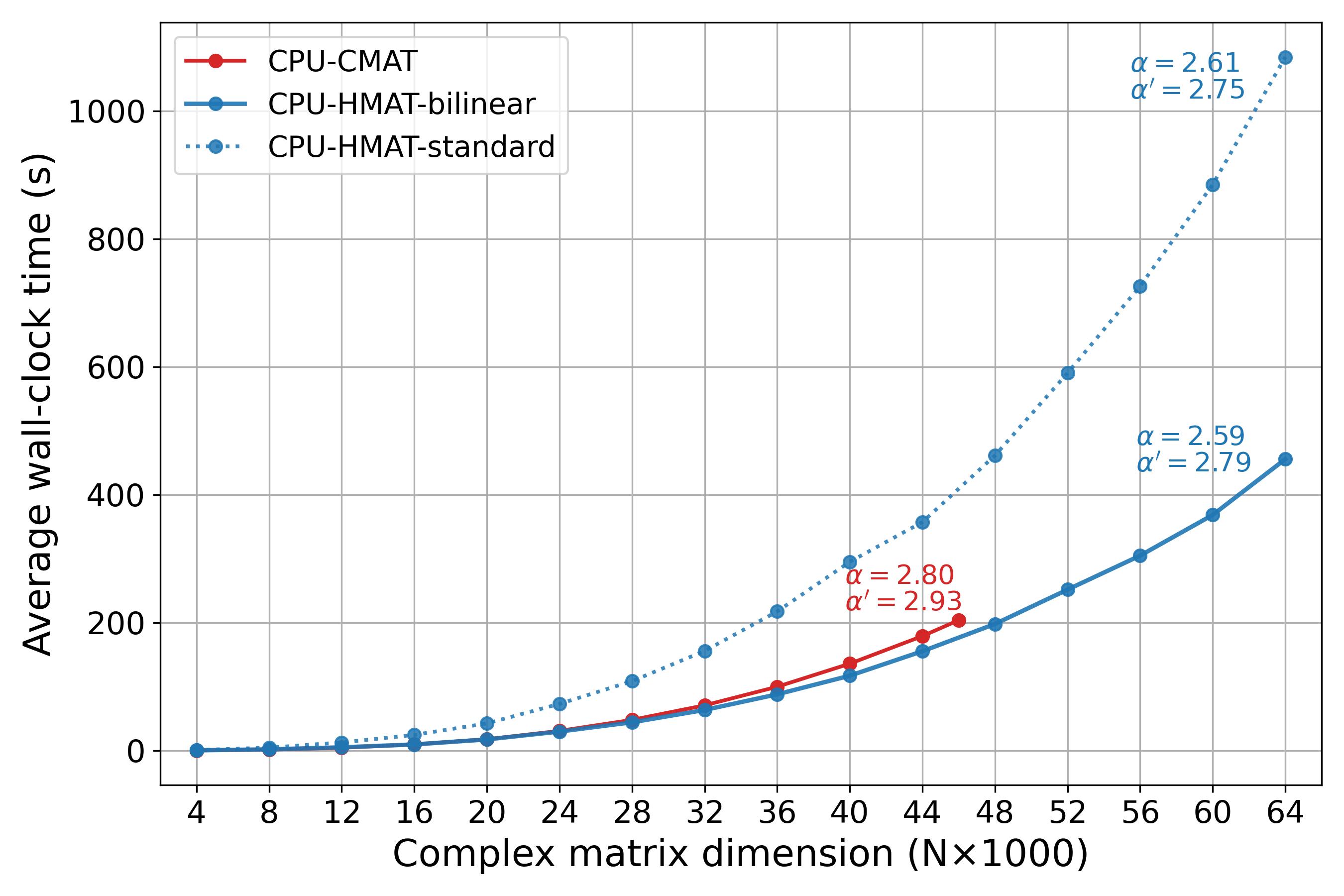}\\
    (a) Pure CPU
    \vspace{0.5em}
    \includegraphics[width=\linewidth]{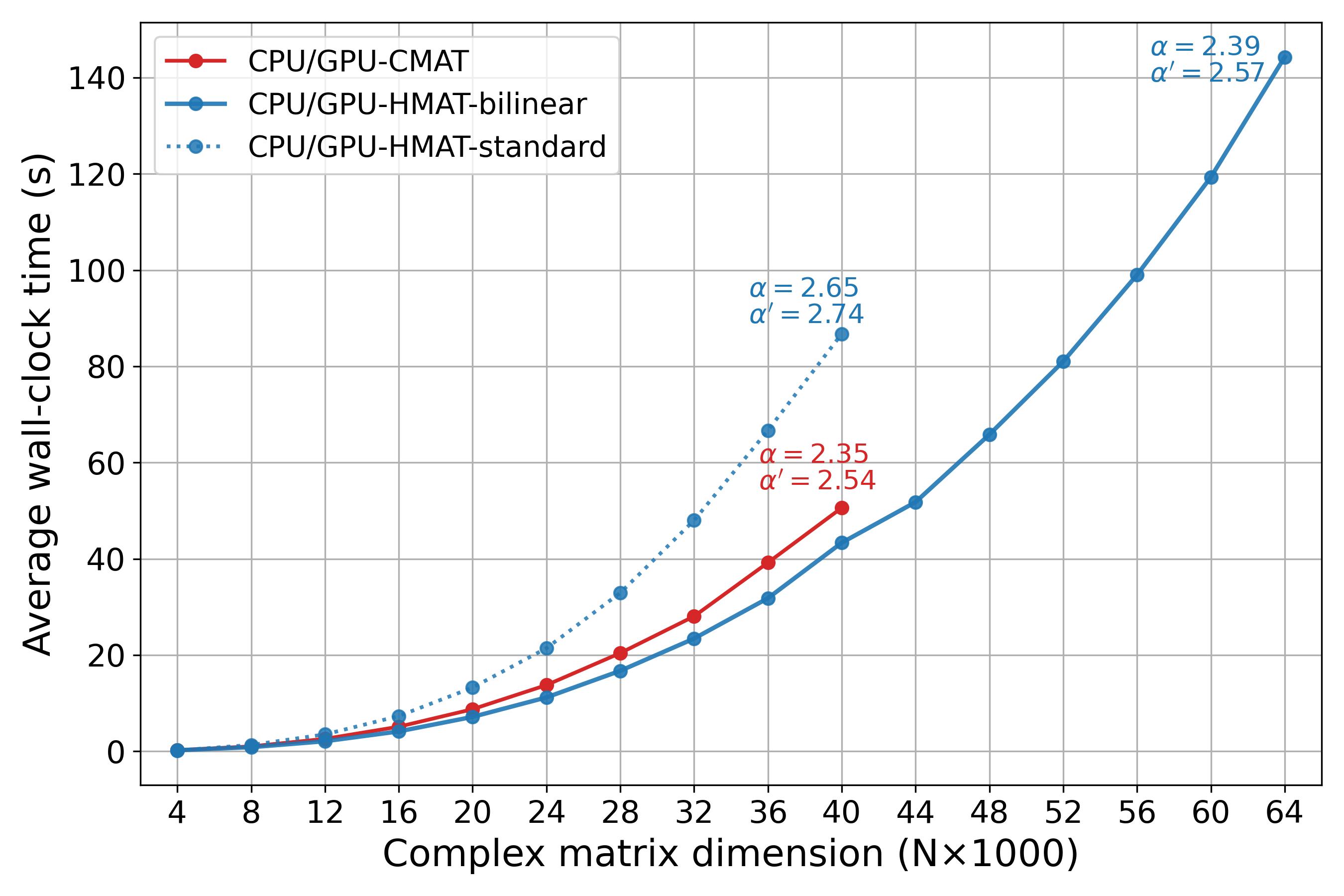}\\
    (b) Hybrid CPU/GPU
  \end{minipage}
  }
\end{minipage}
\hfill
\begin{minipage}{0.48\textwidth}
  \centering
  \fbox{%
  \begin{minipage}{0.97\linewidth}
    \centering
    \textbf{[1,1,1,1,0,0,0,0]}\\[0.5em]
    \includegraphics[width=\linewidth]{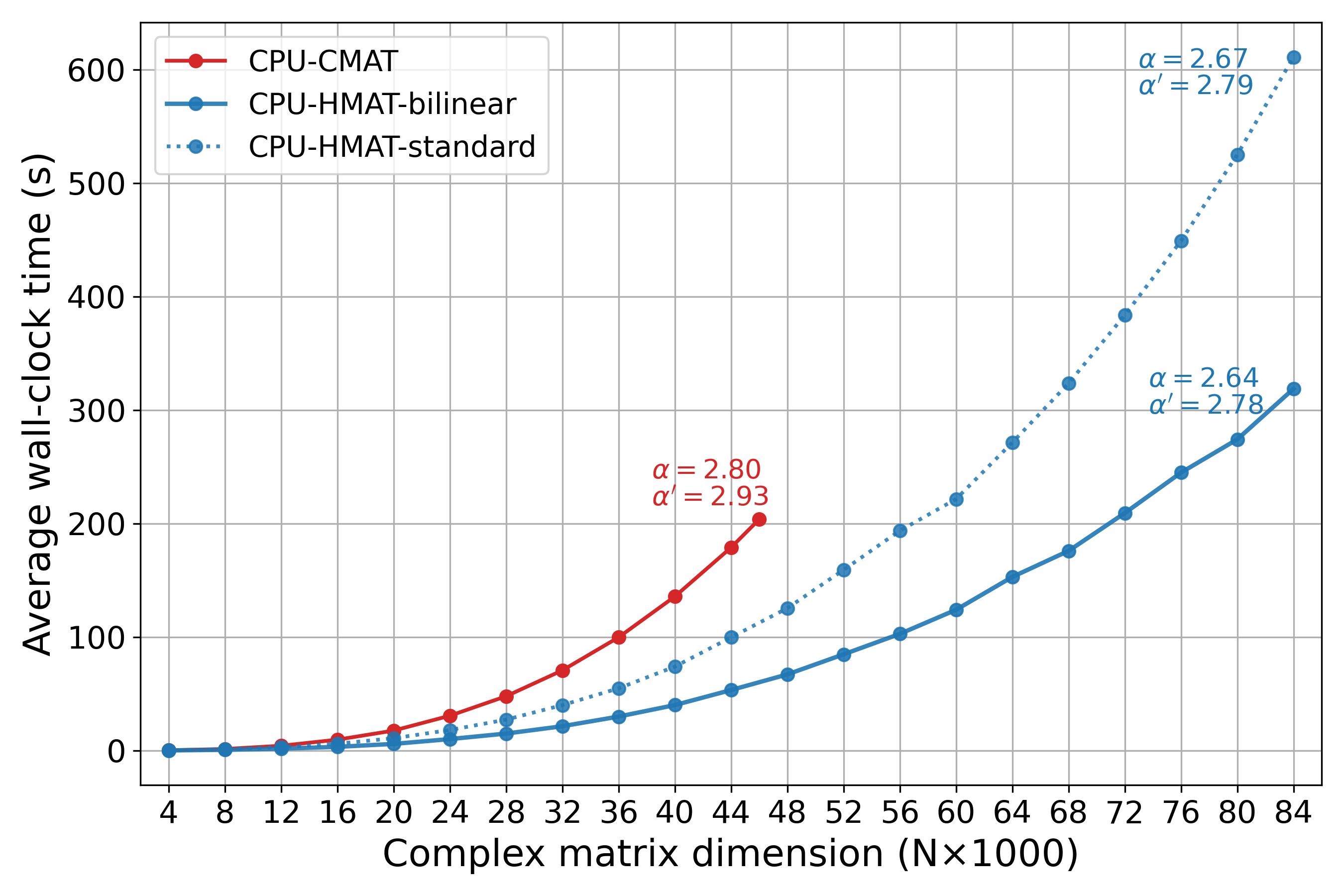}\\
    (c) Pure CPU
    \vspace{0.5em}
    \includegraphics[width=\linewidth]{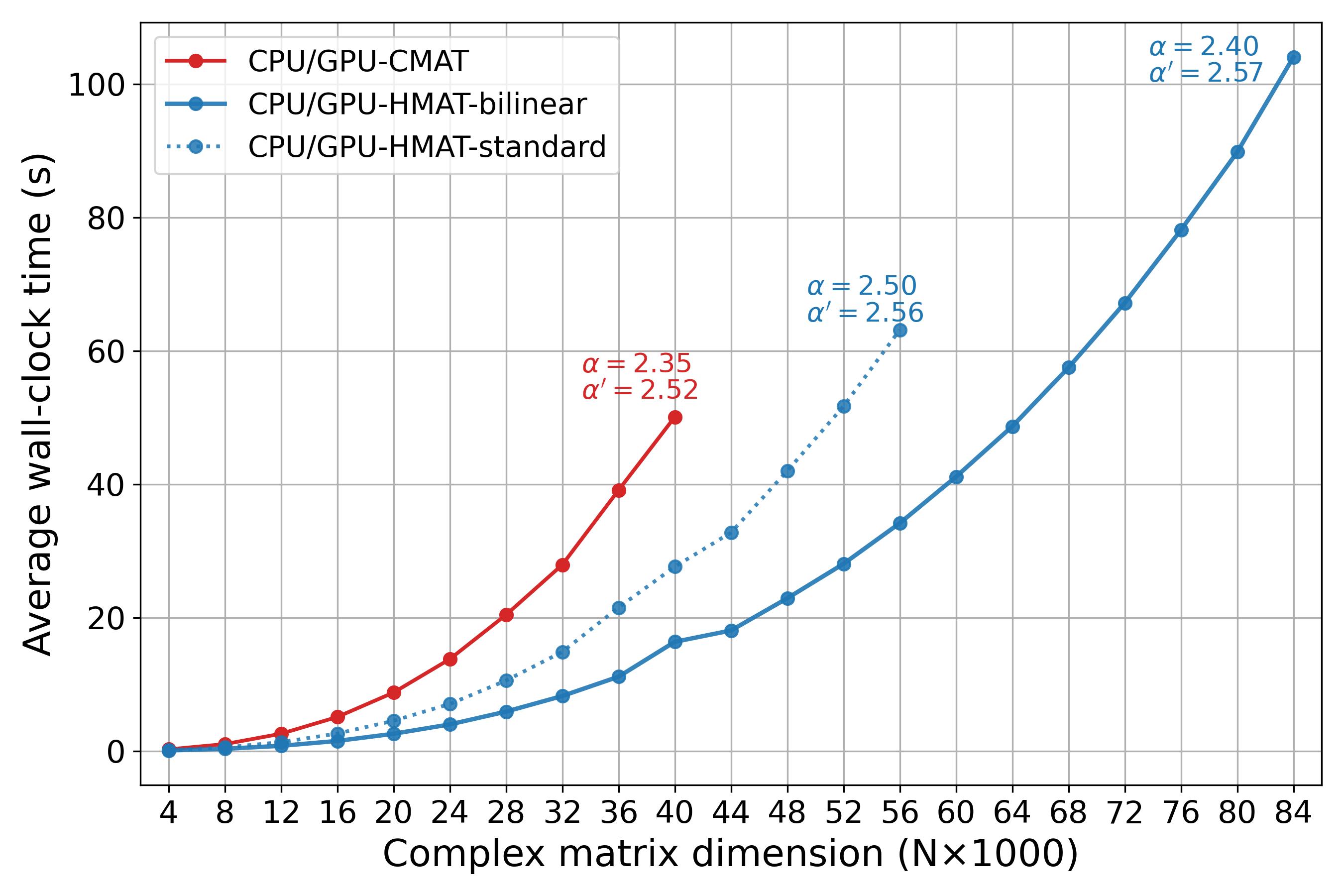}\\
    (d) Hybrid CPU/GPU
  \end{minipage}
  }
\end{minipage}
\caption{Average wall-clock time for complex (\textsc{CMAT}), standard biquaternion (\textsc{HMAT}-standard), and bilinear biquaternion (\textsc{HMAT}-bilinear) matrix--matrix multiplication as a function of the complex matrix dimension $N$. The left and right panels show the fully populated biquaternion ($[1,1,1,1,1,1,1,1]$) and real-quaternion ($[1,1,1,1,0,0,0,0]$) cases, respectively, with panels (a) and (c) corresponding to CPU-only calculations and panels (b) and (d) to hybrid CPU/GPU calculations. Wall-clock times were fitted to a power law, yielding the effective exponents $\alpha$ and $\alpha'$ over the complete range of $N$ and for $N\geq 20000$, respectively. All results were obtained on the \textbf{A100 system} using the \textsc{HMATLIB} library linked with the NVIDIA software libraries specified in Table~\ref{tab:specs}.}
\label{a100-mult}
\end{figure*}

\noindent
Appendix~\ref{app-A:biquaternion-multiplication} reports the results for \textsc{HMATLIB} biquaternion matrix--matrix multiplication on the GH200 [Fig.~\ref{gh200-mult-2}] and A100 [Fig.~\ref{a100-mult}] systems specified in Table~\ref{tab:specs}, complementing the corresponding results and discussion for the H200 system presented in Sec.~\ref{sec:results/multiplication}.

\section{\textsc{HMATLIB} Diagonalization}
\label{app-B:biquaternion-diagonalization}

\begin{figure*}[!t]
    \centering
    \includegraphics[width=0.79\textwidth]{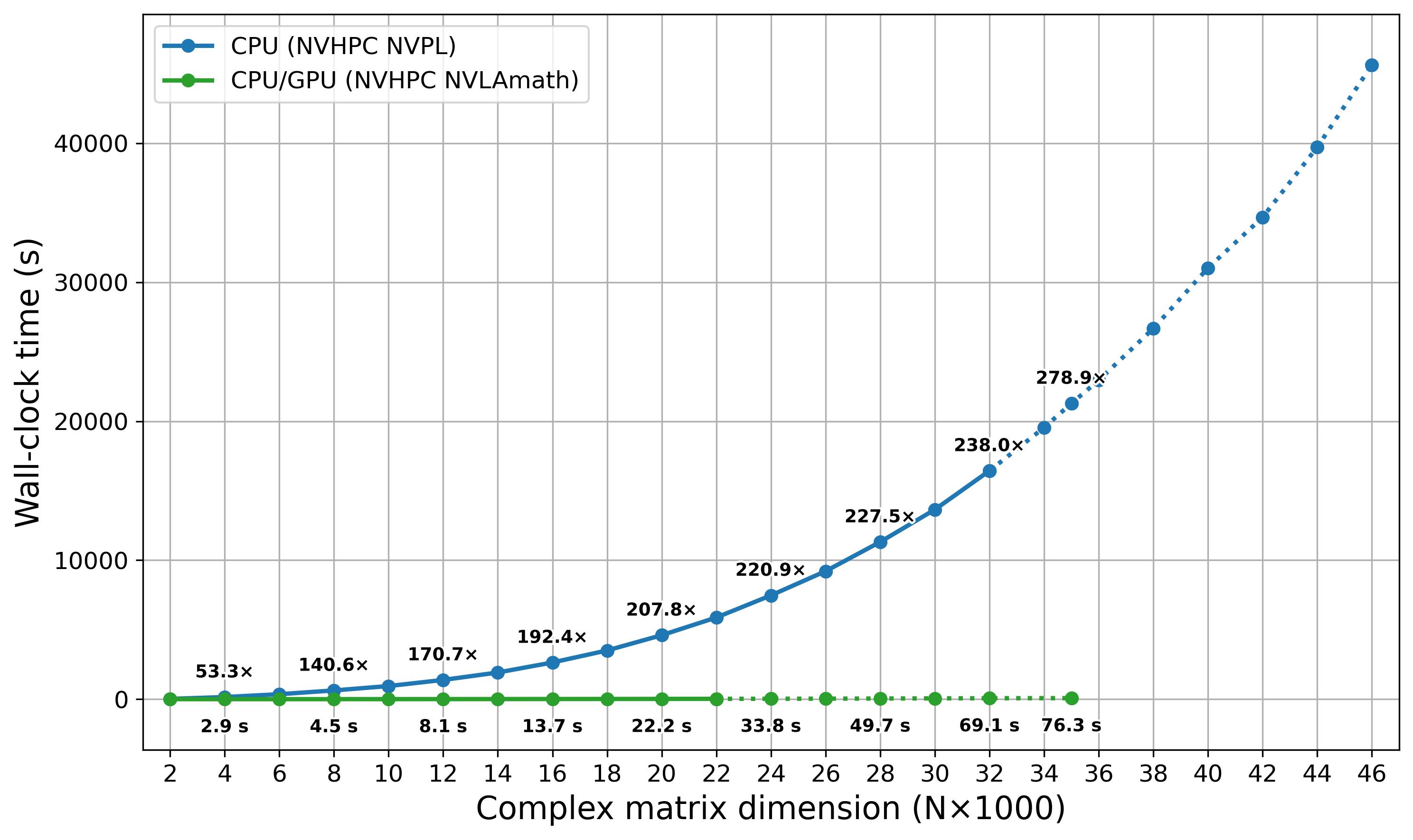}
    \caption{Wall-clock time for \textsc{HMATLIB} diagonalization as a function of the complex matrix dimension $N$. Pure CPU calculations with $144$ threads and NVHPC NVPL backend are compared with hybrid CPU/GPU calculations using NVHPC NVLAmath backend. Values below the green curve report the absolute CPU/GPU execution times in seconds, while the adjacent labels to the CPU curve report the ratio between CPU and CPU/GPU times. The solid curve segments indicate calculations performed with the $32$-bit integer interface while the dotted curve segments correspond to the $64$-bit interface. All calculations were performed on the \textbf{GH200 system} using the software and hardware environments described in Table~\ref{tab:specs}.}
    \label{gh200-diag-qcom8}
\end{figure*}

\begin{figure*}[!t]
    \centering
    \includegraphics[width=0.79\textwidth]{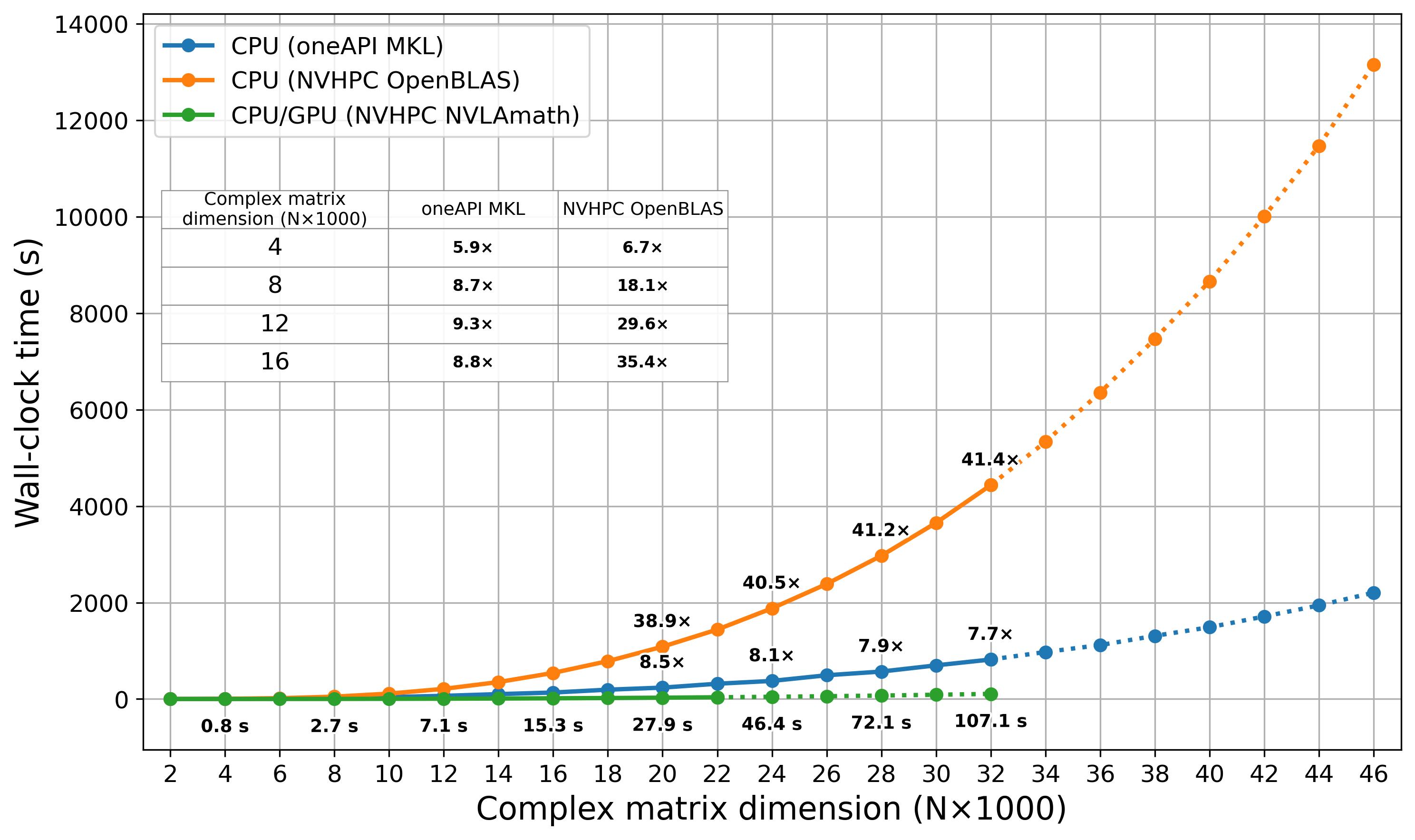}
    \caption{Wall-clock time for \textsc{HMATLIB} diagonalization as a function of the complex matrix dimension $N$. Pure CPU calculations using oneAPI MKL and NVHPC OpenBLAS backends are compared with hybrid CPU/GPU calculations using NVHPC NVLAmath backend. Values below the green curve report the absolute CPU/GPU execution times in seconds, while the adjacent labels to the CPU curves and the inset table report the ratio between CPU and CPU/GPU times. The solid curve segments indicate calculations performed with the $32$-bit integer interface while the dotted curve segments correspond to the $64$-bit interface. All calculations were performed on the \textbf{A100 system} using the software and hardware environments described in Table~\ref{tab:specs}.}
    \label{a100-diag-cpu-gpu-compare-libs}
\end{figure*}

\noindent
Appendix~\ref{app-B:biquaternion-diagonalization} reports the results for \textsc{HMATLIB} diagonalization on the GH200 [Fig.~\ref{gh200-diag-qcom8}] and A100 [Fig.~\ref{a100-diag-cpu-gpu-compare-libs}] systems specified in Table~\ref{tab:specs}, complementing the corresponding results and discussion for the H200 system presented in Sec.~\ref{sec:results/diagonalization}.

For the GH200 CPU diagonalization benchmarks [Fig.~\ref{gh200-diag-qcom8}], an independently installed NVIDIA NVPL 26.5 library was used instead of the version bundled with NVHPC. Its LAPACK implementation provides 64-bit integer interfaces, such as \texttt{zheevd\_64}, within the same library variant, allowing the $32$-bit and $64$-bit eigensolver interfaces to be selected without switching between LP64 and ILP64 libraries. Preliminary thread-count tests were performed using $72$, $144$, $216$, and $288$ OpenMP threads, with $144$ threads providing the most favorable overall performance across the tested matrix dimensions; this configuration was therefore used for the timings reported in Fig.~\ref{gh200-diag-qcom8}.

\end{document}